\documentclass[11pt,aps,amssymb,nofootinbib,amsmath, notitlepage,prd, showkeys]{revtex4-2}
 \providecommand{\keywords}[1]
{
  \small	
  \textbf{\textit{Keywords---}} ;';l';l
}
\usepackage{graphicx}
\usepackage{amsmath}
\usepackage{bm} % bold Greek/math symbols
\usepackage{multirow}
\usepackage{tabularx}
\usepackage{array}
\usepackage{tensor} % tensor indices
\usepackage{slashed} % Dirac slash
\usepackage{relsize} % rescalable math 
\usepackage{soul} % letter spacing, underlining
\usepackage{yfonts} % Gothic fonts
\usepackage{float}
\usepackage{orcidlink}
\usepackage{hyperref}
\def\bal#1\gal{\begin{align}#1\end{align}}

\begin{document}
\title{Nuclear Transformations of Atoms under the Influence of Acoustic Oscillations on Water}
\author{R. N. Balasanyan$^{1}$\orcidlink{0009-0002-5026-6854}}
\author{S. G. Minasyan$^{2}$}
\author{I. G. Grigoryan$^{1}$}
\thanks{Corresponding author}
\email{irina.g.grigoryan@gmail.com}
\author{V. S. Arakelyan$^{1}$}
\author{R. B. Kostanyan$^{1}$}

\affiliation{$^{1}$Institute for Physical Research, National Academy of Sciences of Armenia, Gitavan-2, 0204 Ashtarak, Republic of Armenia}
\affiliation{$^{2}$Institute of Chemical Physics after A.B. Nalbandyan, National Academy of Sciences of Armenia, 5/2, Paruir Sevak St., Yerevan 0014, Republic of Armenia}
        \begin{abstract}
The results of studies on the properties of ordinary and heavy water subjected to sharp  mechanical impacts at acoustic repetition frequency are presented. Experimental evidence for the phenomenon of acoustically induced nuclear processes in water is provided, supported by direct measurements of the presence of radiation emission and the subsequent formation of new elements, which cannot be explained by chemical reactions. The complex nature of the influence of mechanical oscillations on the concentration changes of stable isotopes of chemical elements such as \textbf{Ti, B, Na, Mg,} and \textbf{Li} in water is demonstrated. The cause of surface erosion of metal structures during cavitation in water is explained through the formation of fluorine and, consequently, the creation of aggressive \textbf{HF} acid molecules. A mechanism for the occurrence of sonoluminescence under sharp acoustic impact on water is proposed.
\end{abstract}

        \maketitle
        \section{Introduction}\label{sec:intro}
Numerous reports in the scientific literature describe the formation of new atomic nuclei under intense mechanical impacts on deuterium-containing substances. Among the various phenomena, nuclear transformations caused by acoustic impacts on heavy water are particularly significant. Studies  \cite{1,2,3,4,5,6,7,8,9,10,11,12,13,14,15,16} have shown that intense mechanical impacts on deuterium-containing solids result in neutron emission. This emission is also observed during the deuteration of titanium and palladium electrodes during electrolysis in heavy water (D$_2$O)\cite{3}. Study \cite{6} found a direct correlation between neutron burst intensity and acoustic emission signals associated with the cracking of deuterium-saturated titanium and palladium samples. This process is accompanied by visible light emission, radio waves, and X-ray radiation detected during cavitation in water \cite{4}. These findings suggest the formation of strong electric fields (10$^6$--10$^7$ V/cm), similar to those observed in dielectric crystal fractures \cite{5}.  
Study \cite{3} found no excess neutron emission during cavitation in ordinary water (H$_2$O) or heavy water (D$_2$O) when using a fresh titanium vibrator. However, after 18-20 hours of use, signs of erosion appeared on the end surface of the titanium vibrator submerged in (D$_2$O). After this, the number of neutrons emitted during cavitation significantly exceeded natural background levels. This effect disappeared away from resonance or after cavitation stopped. A similar experiment using pure water showed no such effect \cite{3}.

The reviewed studies link neutron emission to low-temperature nuclear fusion in heavy water through DD-reactions. This assumption is based on the hypothesis that temperatures inside cavitation bubbles can reach 10$^8$K, with pressures of 10$^5$--10$^6$  atm \cite{6}.

Furthermore, studies \cite{7,8,9,10,11,12,13,14,15,16} suggest that the cavitation of bubbles generates high temperatures over a short time, producing a phenomenon called sonoluminescence. This is thought to be accompanied by nuclear fusion within collapsing gas bubbles. Some researchers, however, argue that the high temperature inside the bubbles is a result of nuclear fusion, rather than its cause. In addition, during sonoluminescence, the intensity of the glow from bubbles depends significantly on the hydrogen-to-deuterium ratio in the liquid being studied.

In these and other studies, researchers convincingly demonstrate neutron emission under ultrasonic radiation, linking it to low-temperature nuclear fusion. However, some of the arguments and proposed mechanisms are not fully convincing. Studies \cite{17, 18} observed radiation emission in ordinary water under sharp mechanical vibrations at acoustic frequencies, even in the absence of cavitation and sonoluminescence. The mechanism of nuclear processes in ordinary water under mechanical impacts with acoustic frequency is convincingly presented in \cite{17, 18}.
Study \cite{19} reports the results of experimental research on nuclear processes in heavy water under normal conditions. When heavy water was exposed to mechanical pulses with sharp fronts at acoustic frequencies, gamma radiation was detected, without cavitation or sonoluminescence. Additionally, significant changes in impurity concentrations in the liquids were observed, both quantitatively and qualitatively.
Clearly, these phenomena are of great interest to both academic science and applied fields like materials science.
        \section{Methodology of Acoustic Vibrations Impact on Water}\label{sec:setUp}
\begin{figure}
    \centering
    \includegraphics[width=4in]{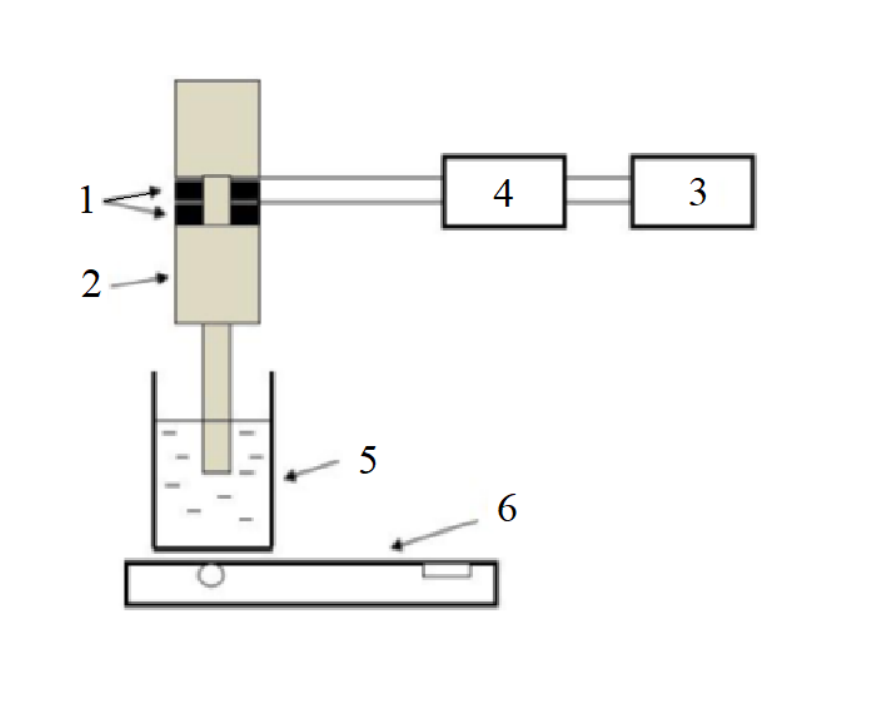} 
    \caption{Schematic diagram of the experimental setup.}
    \label{fig:1}
\end{figure}

In studies \cite{17,18,19}, radiation emission was observed in ordinary water under the influence of mechanical vibrations at an acoustic repetition frequency, without cavitation or sonoluminescence phenomena. The basic schematic of the experimental setup for mechanical impact on ordinary water is shown in Fig.~\ref{fig:1}.

Piezoelectric washers (1) with a titanium concentrator (2), tuned to a resonant frequency of 19.3 kHz, served as the source of ultrasonic emissions. At resonance, the sound level reached 120 dB. The piezoelectric washers were powered by a generator (3) and an amplifier (4).
When water (5) was exposed to ultrasound with the specified parameters, no radiation emissions were detected by the \textbf{DKS-04} dosimeter (6) or the \textbf{RUP-1} radiometer. However, when the shape of the electrical signal supplied to the piezoelectric ceramics was changed from sinusoidal to rectangular, the devices registered \boldsymbol{$\gamma$}-radiation. Radiation was also detected away from acoustic resonance at sound levels of 75 dB or lower, across both audible and infrasonic frequencies. In the absence of contact between the titanium concentrator and water, \boldsymbol{$\gamma$}-radiation signals were not detected. 

To validate the measurement results, studies of the acoustic impact on ordinary water were conducted using a precision low-background \boldsymbol{$\gamma$}-spectrometer manufactured by Camverra. Additionally, the chemical composition of the water was analyzed at various durations of ultrasonic exposure. Water composition measurements were performed using a \textbf{Nex ION 9000} mass spectrometer from \textbf{Perkin Elmer and a Dionex ICS-1000} device.

        \section{Detection of Nuclear Transformations During Acoustic Exposure to Water}
\begin{figure}[H]
    \centering
    \includegraphics[width=5in]{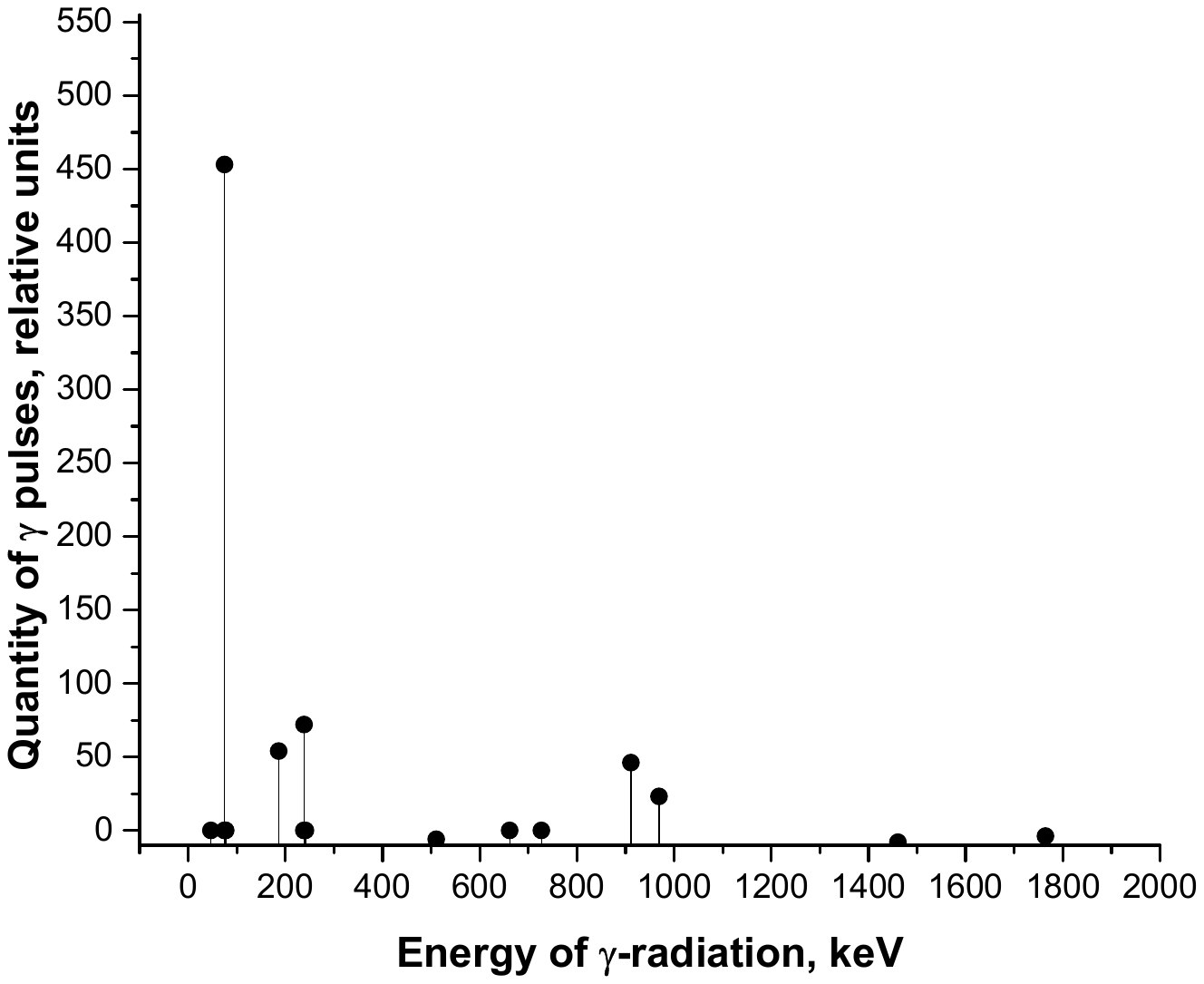} 
    \caption{Spectrum of \boldsymbol{$\gamma$}-radiation from the setup during contact action of the acoustic emitter on ordinary water.
    }
    \label{fig:2}
\end{figure}
Measurements on the \boldsymbol{$\gamma$}-spectrometer were carried out with a one-hour accumulation time. The background radiation level was measured both when the acoustic emitter was operating without contact with water and when the titanium tip of the ultrasonic emitter was submerged in water. Fig.~\ref{fig:2} shows the \boldsymbol{$\gamma$}-radiation spectrum from the system during operation, with the background level subtracted from measurements taken without water.

Study~\cite{17} confirms the presence of \boldsymbol{$\gamma$}-radiation during the interaction between an acoustic field and water, suggesting the occurrence of potential nuclear processes. This radiation has been demonstrated to be unrelated to cavitation in water under ultrasonic influence, as it is observed even under non-resonant conditions and without sonoluminescence.
Significant changes in the properties of water are detected when subjected to such acoustic exposure. Specifically, it was found that sound vibrations with steep fronts lead to a substantial increase in the electrical conductivity of ordinary water. For instance, exposure of 100 mL of water to a 30 kHz frequency for 10 hours results in more than a twofold increase in the liquid's electrical conductivity.

\begin{figure}
    \centering
    \includegraphics[width=1.7in]{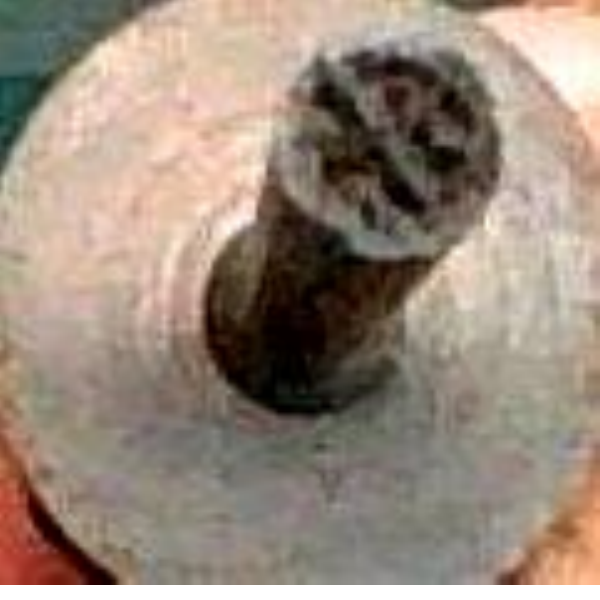} 
    \caption{Image of the etched surface at the end of the titanium concentrator of the acoustic emitter.}
    \label{fig:3}
\end{figure}

Additionally, after several hours of exposure to these vibrations, erosion marks begin to appear on the end surface of the titanium concentrator, which become more pronounced with continued vibrational impact. Fig.~\ref{fig:3} shows an image of the etched surface of the titanium concentrator's end as a result of the acoustic emitter's activity.
A similar phenomenon of etching the end of an acoustic emitter is described in \cite{3}. The authors of that study associate the observed phenomenon with the cavitation process in heavy water and note that such a phenomenon does not occur in ordinary water. However, the phenomenon described above, as reported in \cite{17,18,19}, was observed in ordinary water without the cavitation process.
\begin{figure}
\centering
    \includegraphics[width=3in]{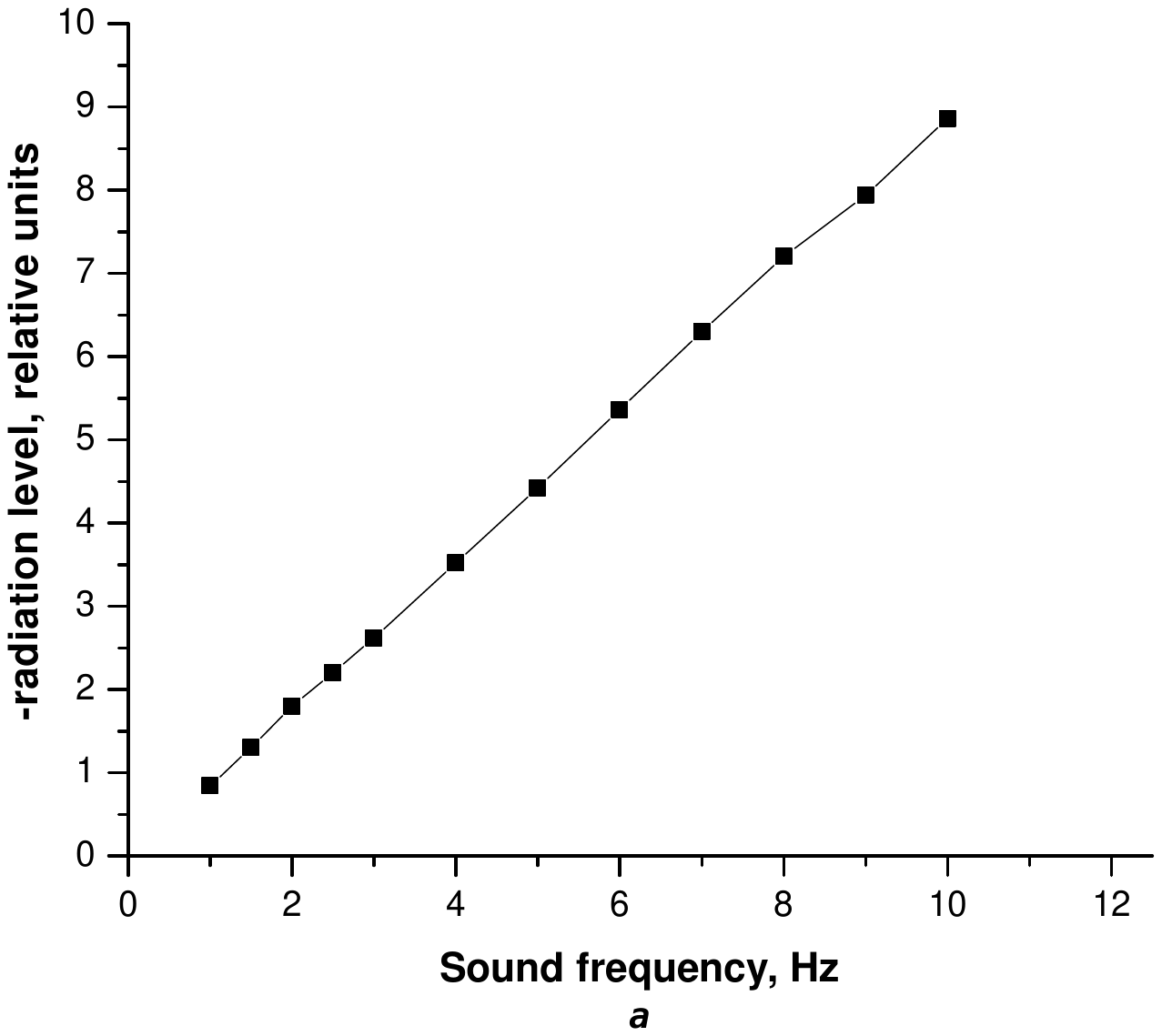} 
    \includegraphics[width=3in]{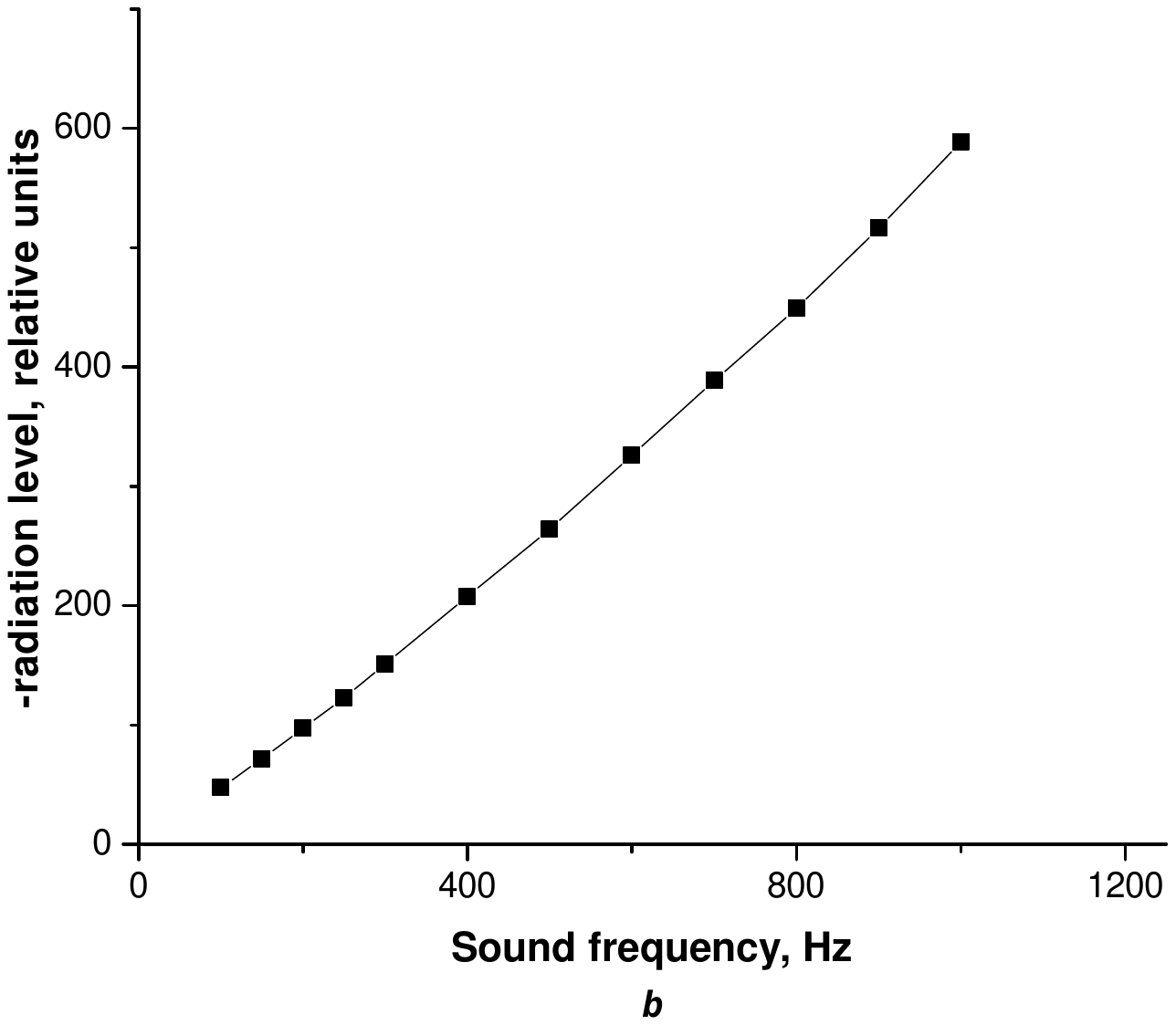} \\
    \caption{Dependence of \boldsymbol{$\gamma$}-radiation level in relative units on sound frequency in the range of 1 to 10 Hz (a) and 100 to 1000 Hz (b).
    }
    \label{fig:4}
\end{figure}

Fig.~\ref{fig:4} illustrate the dependence of the \boldsymbol{$\gamma$}-radiation level on the sound frequency within the infrasonic range from 1 to 10 Hz (Fig.~\ref{fig:4}a) and from 100 to 1000 Hz Fig.~\ref{fig:4}b). As shown by the graphs, a predominantly linear increase in the \boldsymbol{$\gamma$}-radiation level is observed. This type of dependence can be interpreted as an increase in the number of sound pulses acting on the water per unit of time. 
\begin{figure}
    \centering
    \includegraphics[width=4.5in]{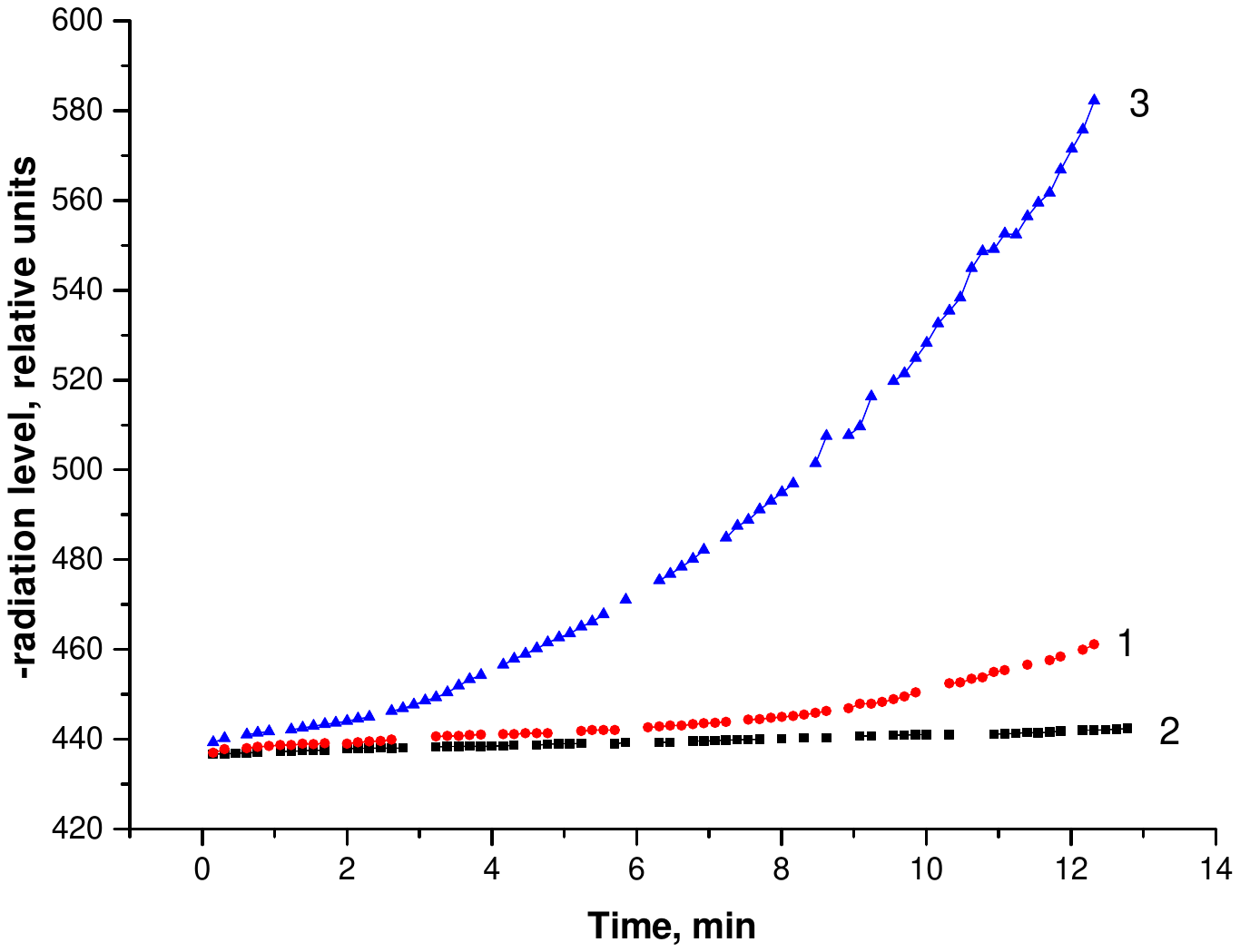} 
    \caption{Dependence of $\gamma$-radiation level in relative units on time, min for ordinary water (1), alkaline aqueous solution with pH $\approx$ 10 (2), and acidic aqueous solution with pH $\approx$ 1 (3).}
    \label{fig:5}
\end{figure}

Another noteworthy result is the change in the \boldsymbol{$\gamma$}-radiation level over time at the same sound frequency.
Fig.~\ref{fig:5} shows the dependence of the \boldsymbol{$\gamma$}-radiation level in relative units on time, measured in liquids with varying pH levels. In Fig.~\ref{fig:5}, Curve 1 represents measurements from ordinary water; Curve 2 reflects results obtained from an aqueous \textbf{NaOH} solution with pH$\approx$10; and Curve 3 corresponds to measurements conducted in an aqueous \textbf{HCl} solution (pH$\approx$1). The graphs indicate that a certain cumulative process is occurring, which is less pronounced in alkaline conditions compared to acidic ones. The level of \boldsymbol{$\gamma$}-radiation is higher in solutions with a greater concentration of hydrogen ions. It is possible that the increased hydrogen ion concentration contributes to the formation of double electric layers.

To confirm changes in the concentration of chemical elements, potentially indicating nuclear processes in the studied water samples, measurements were conducted using a \textbf{Perkin Elmer ICP MS ELAN 9000} mass spectrometer. Distilled water samples were subjected to abrupt mechanical action at a frequency of 25 kHz for exposure periods of three and six hours. Given that the concentrator is made of titanium and may alter the composition of the water upon contact, the entire volume of the source water was pre-conditioned by being in contact with the concentrator for eight hours. The mass spectrometer measurements revealed the presence of a significant number of chemical elements in the water samples, with their concentrations having changed depending on the conditions of ultrasonic exposure.
\begin{table}[H]
\centering
\scriptsize 
\begin{tabular}{|>{\bfseries\centering\arraybackslash}m{3cm}|>{\centering\arraybackslash}m{3cm}|>{\centering\arraybackslash}m{3cm}|>{\centering\arraybackslash}m{3cm}|>{\centering\arraybackslash}m{3cm}|}
\hline
\multirow{2}{*}{\textbf{Chemical Element}} & \multirow{2}{*}{\textbf{Atomic Mass (amu)}} & \multicolumn{3}{c|}{\textbf{Impurity Concentration (mg/L)}} \\ 
\cline{3-5} 
 &  & \textbf{Initial Sample} & \textbf{After 3 Hours} & \textbf{After 6 Hours} \\ 
\hline
Li  & 7   & 0.000050  & 0.000141  & 0.000277  \\ \hline
Be  & 9   & 0.000005  & 0.000005  & 0.000006  \\ \hline
B   & 11  & 0.017771  & 0.018707  & 0.020833  \\ \hline
Na  & 23  & 0.000322  & 0.001200  & 0.001314  \\ \hline
Mg  & 24  & 0.008085  & 0.009921  & 0.042501  \\ \hline
Al  & 27  & 0.004729  & 0.005474  & 0.004372  \\ \hline
K   & 39  & 0.004399  & 0.005387  & 0.006079  \\ \hline
Ca  & 43  & 0.013358  & 0.014885  & 0.027868  \\ \hline
Ti  & 48  & 0.0006079 & 0.0006595 & 0.001363  \\ \hline
V   & 51  & 0.000146  & 0.000173  & 0.000305  \\ \hline
Cr  & 52  & 0.004179  & 0.004408  & 0.004783  \\ \hline
Mn  & 55  & 0.001041  & 0.001004  & 0.000971  \\ \hline
Fe  & 54  & 0.015043  & 0.014361  & 0.014648  \\ \hline
N   & 59  & 0.000021  & 0.000021  & 0.000018  \\ \hline
Ni  & 60  & 0.000048  & 0.000157  & 0.000162  \\ \hline
Cu  & 63  & 0.000245  & 0.000847  & 0.000899  \\ \hline
Zn  & 66  & 0.002575  & 0.006509  & 0.005330  \\ \hline
Ga  & 69  & 0.000456  & 0.000484  & 0.000405  \\ \hline
Ge  & 74  & 0.000015  & 0.000016  & 0.000017  \\ \hline
As  & 75  & 0.000064  & 0.000064  & 0.000069  \\ \hline
Se  & 76  & 0.007468  & 0.007815  & 0.008043  \\ \hline
Rb  & 85  & 0.000165  & 0.000173  & 0.000218  \\ \hline
Sr  & 88  & 0.001158  & 0.001318  & 0.002385  \\ \hline
Pd  & 106 & 0.000169  & 0.000106  & 0.000142  \\ \hline
Mo  & 98  & 0.000041  & 0.000052  & 0.000065  \\ \hline
Ru  & 102 & 0.000008  & 0.000008  & 0.000009  \\ \hline
Rh  & 103 & 0.000616  & 0.000675  & 0.000663  \\ \hline
Ag  & 107 & 0.000006  & 0.000008  & 0.000043  \\ \hline
Cd  & 111 & 0.000009  & 0.000067  & 0.000019  \\ \hline
Sn  & 118 & 0.000028  & 0.000028  & 0.000032  \\ \hline
Sb  & 121 & 0.005497  & 0.006442  & 0.021572  \\ \hline
Cs  & 133 & 0.000004  & 0.000007  & 0.000009  \\ \hline
Ba  & 138 & 0.007607  & 0.008250  & 0.007038  \\ \hline
W   & 184 & 0.000007  & 0.000012  & 0.000012  \\ \hline
Re  & 187 & 0.000000  & 0.000001  & 0.000000  \\ \hline
Ir  & 193 & 0.000001  & 0.000001  & 0.000001  \\ \hline
Pt  & 195 & 0.000002  & 0.000002  & 0.000002  \\ \hline
Au  & 197 & 0.000003  & 0.000003  & 0.000003  \\ \hline
Tl  & 205 & 0.000001  & 0.000001  & 0.000001  \\ \hline
Pb  & 208 & 0.000449  & 0.000171  & 0.000177  \\ \hline
U   & 238 & 0.000002  & 0.000001  & 0.000002  \\ \hline
% Add more rows as needed
\end{tabular}
% \end{adjustbox}
\caption{Chemical elements detected in water samples after exposure to acoustic waves at different time intervals (0, 3, and 6 hours)}
\label{tab:table1}
\end{table}
Table ~\ref{tab:table1} lists the chemical elements detected in the water samples subjected to acoustic waves with exposure times of 0, 3, and 6 hours.
It follows from the table that under the given conditions of exposure to distilled water, the concentration of some elements, such as $Li, B, Na, Mg, K, Ca, V, Cr, Ni, Cu, Ge, Se, Rb, Sr, Mo, Ag, Sb$, and $Cs$, increases with longer ultrasonic exposure. The concentrations of $Mn$ and $N$ decrease with increasing exposure time. The concentrations of $Be, Ru, Re, Ir, Pt, Au, Tl$, and $U$ remain almost unchanged within the applied exposure conditions and measurement accuracy. More complex changes in the concentrations of certain elements were also observed under ultrasound exposure. 

\begin{figure}[H]
    \centering
    \includegraphics[width=4.5in]{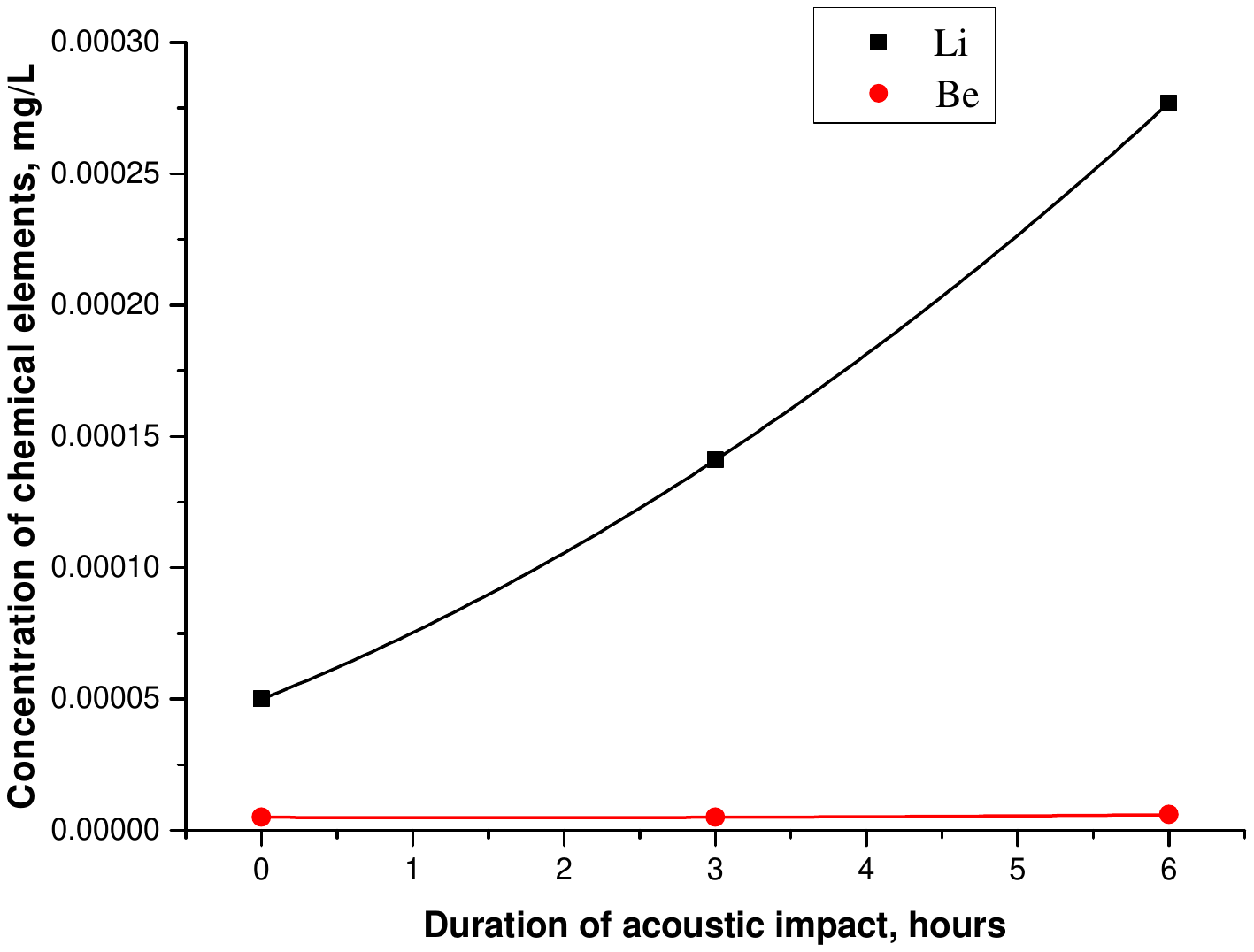} 
    \caption{Dependence of the concentration changes of lithium and beryllium isotopes (mg/L) on the impact duration of acoustic oscillations on distilled water.}
    \label{fig:6}
\end{figure}

For illustration, Fig.~\ref{fig:6} shows the variation in the concentrations of \textbf{lithium}-7 ($_{3}{}Li^7$) and \textbf{beryllium}-9 ($_{4}{}Li^9$) isotopes with increasing exposure time to oscillations on distilled water. The graph shows a clear increase in the concentration of lithium, while the concentration of beryllium remains largely unaffected by the duration of ultrasonic exposure.
Research indicates that sharp acoustic oscillations applied to water lead to measurable changes in the concentrations of impurity ionites. Table ~\ref{tab:table2} presents the results of the concentration measurements for selected ionites present in the water, obtained using the \textbf{Dionex ICS-1000} system.
It can be seen from the table ~\ref{tab:table2} that as the exposure duration increases to six hours, the concentration of fluorine in the studied samples increases.
\begin{table}[H]
\centering
\scriptsize
\begin{tabular}{|>{\bfseries\centering\arraybackslash}m{3cm}|>{\centering\arraybackslash}m{3cm}|>{\centering\arraybackslash}m{3cm}|>{\centering\arraybackslash}m{3cm}|>{\centering\arraybackslash}m{3cm}|}
\hline
\multirow{2}{*}{\textbf{Chemical Element}} & \multirow{2}{*}{\textbf{Atomic Mass (amu)}} & \multicolumn{3}{c|}{\textbf{Impurity Concentration (mg/L)}} \\ 
\cline{3-5} 
 &  & \textbf{Initial Sample} & \textbf{After 3 Hours} & \textbf{After 6 Hours} \\ 
\hline
N       & 14          & 0.0350  & 0.0000  & 0.0456  \\ \hline
F       & 19          & 0.0057  & 0.0061  & 0.0074  \\ \hline
S       & 32          & 0.1017  & 0.0156  & 0.0387  \\ \hline
Cl      & 35          & 0.0000  & 0.0341  & 0.0329  \\ \hline
\end{tabular}
\caption{Concentration measurements of selected ionites in water after exposure to sharp acoustic oscillations at 0, 3, and 6 hours}
\label{tab:table2}
\end{table}

         \section{Physical mechanisms of nuclear processes in water induced by acoustic vibrations}
It is well-established that nuclear fusion requires extreme conditions, such as high temperature (T $\approx$10$^8$ K), high pressure (P $\approx$ 10$^6$ atm), and the fulfillment of the Lawson criterion: $\boldsymbol{n\tau>}\boldsymbol{10^{14} \, \text{s/cm}^3}$, where $n$ is the density of high-temperature plasma, and $\tau$ is the confinement time within the system. In the described conditions of ultrasound exposure to ordinary water, neither high pressure nor high temperature was present. Additionally, cavitation and sonoluminescence were not observed during the exposure; however, sensors recorded the presence of gamma radiation (\boldsymbol{$\gamma$}-radiation).

Another possible condition for the occurrence of nuclear processes is the presence of thermal neutrons. Under typical conditions, when there is no way to overcome the Coulomb barrier of nuclei, nuclear reactions occur through the ($n$,\boldsymbol{$\gamma$}) channel. When an ultrasonic titanium concentrator is in contact with water, a double electric layer forms at the boundary between the water and titanium. This layer is characterized by an intense electric field with a strength of approximately 10$^6$--10$^7$ V/cm. The titanium boundary layer acquires a negative charge due to thermal electrons, while the water layer near the titanium becomes positively charged due to dissociated hydrogen ions.

External influences can alter the characteristics of this double electric layer. A sharp mechanical impact from the titanium rod on the liquid, directed perpendicular to the double electric layer, can disrupt the integrity of the layers, potentially leading to collisions between hydrogen ions and electrons at the surface layer of the titanium concentrator. By analogy with K-capture, this process may result in the formation of neutrons through the following reaction channel:
\[
\text{H}^+ + e^- \rightarrow n^* + \nu \tag{1} \label{eq:1}
\]
The neutron ($n^*$) formed via this mechanism possesses a mass defect, meaning it is a quasi-neutron. To compensate for the missing mass, the quasi-neutron must engage in an energy interaction with the nearest nucleus, transforming it into an isotope of that atom. Thus, a nuclear reaction of the ($n$, \boldsymbol{$\gamma$}) type is possible, which may account for the relatively high level of \boldsymbol{$\gamma$}-radiation observed during sharp acoustic vibrations applied to water.

The generated neutrons can interact not only with titanium atoms but also with the nuclei of hydrogen atoms, converting them into hydrogen isotopes—deuterium and tritium—which are not \boldsymbol{$\gamma$}-radioactive. Therefore, it can be hypothesized that the \boldsymbol{$\gamma$}-activity is also linked to the transformation of oxygen nuclei present in water into fluorine nuclei; specifically, the conversion of an oxygen nucleus into a fluorine nucleus with the emission of a \boldsymbol{$\gamma$}-quantum: \(\text{O}(n,\boldsymbol{\gamma})\text{F}\).  The half-life of this transformation is 27 seconds \cite{20}. Schematically, this process can be represented as follows:
\[
_{8}{}\text{O}^{18} + n^* \rightarrow (_{8}{}\text{O}^{19})^* \rightarrow _{9}{}\text{F}^{19} + \beta^- + \tilde{\nu} + \gamma \tag{2} \label{eq:2}
\]
The Oxygen atom $_{8}{}\text{O}^{18}$ is a stable isotope of oxygen, constituting 0.2\% of naturally occurring oxygen isotopes. When it captures a neutron, it forms the isotope $_{8}{}\text{O}^{19}$, which is a \boldsymbol{$\beta^-$}-active, unstable isotope and decays into the stable isotope of fluorine. As seen in Table~\ref{tab:table2}, the concentration of fluorine in water increases with the duration of ultrasonic exposure. Fluorine is a chemically active element with a strong affinity for hydrogen, enabling it to remove hydrogen even from robust compounds like water.

When oxygen in the water molecule is converted to fluorine, one of the hydrogen ions must leave the former water molecule, transforming it into hydrofluoric acid (HF), a chemically aggressive substance. The detached hydrogen ion will enhance the quasi-neutron formation process, which can lead to an increase in the number of nuclear transformation events from oxygen to fluorine, thereby increasing the intensity of emitted \boldsymbol{$\gamma$}-quanta. This provides a possible explanation for the mechanism of increasing \boldsymbol{$\gamma$}-radiation levels over time, as shown in Fig.~\ref{fig:5}.
The formation of hydrofluoric acid in water could also explain the etching phenomenon on the tip of the ultrasonic sensor, as depicted in Fig.~\ref{fig:3}. From the results of experimental studies, there appears to be no correlation between the level of \boldsymbol{$\gamma$}-radiation and the phenomenon of sonoluminescence. In studies \cite{7,8,9,10,11,12,13,14,15,16}, the mechanism of light flashes under powerful ultrasonic radiation (sonoluminescence) is associated with the luminescence of \textbf{OH} radicals present in water. This approach conflicts with the continuous optical spectrum of emission observed in the blue-violet region of the spectrum.

The phenomenon of sonoluminescence occurs in parallel with the cavitation process. Cavitation bubbles form double electric layers between their inner and outer surfaces. During the rarefaction phase of the acoustic field in water, dissolved gases and water vapors create two regions: an internal, rarefied bubble cavity and the surrounding water environment. A double electric layer forms at the boundary between these regions. During the compression phase, the bubbles collapse, resulting in the discharge of the double electric layer, which explains the appearance of a continuous spectrum and the spectral emission during sonoluminescence.  

The complex changes in the concentrations of chemical elements (Table~\ref{tab:table1} and Table~\ref{tab:table2}) in water after sharp mechanical impacts can be explained by the following reasons:
\begin{itemize}
    \item Differences in the cross-sections of nuclear reactions of the type ($n$, \boldsymbol{$\gamma$});
    \item The half-life of a radioactive isotope formed after neutron capture;
    \item The proximity of the element to the titanium concentrator, near which quasi-neutrons are generated according to reaction (\ref{eq:1}).
\end{itemize}

From Fig.~\ref{fig:6}, it can be observed that, within the measurement accuracy, no significant change in the concentration of beryllium ($Be$) occurs, even though quasi-neutron capture by the $_{4}{}Be^9$ nucleus and subsequent nuclear transformation could take place via the following channel:
\[
_{4}{}\text{Be}^{9} + n^* \rightarrow (_{4}{}\text{Be}^{10})^* \rightarrow _{5}{}\text{B}^{10} + \beta^- + \tilde{\nu} + \gamma \tag{3} \label{eq:3}
\]
The half-life of the radioisotope \textbf{beryllium-10} ($_{4}{}Be^{10}$) is 1.6 $\cdot$$10^6$ years~\cite{20}. Compared to this long half-life, the 6-hour exposure time is negligible, which is why no noticeable changes in beryllium concentration are observed in the water under acoustic vibrations.

An analysis of a potential increase in beryllium concentration via the ($n$, \boldsymbol{$\gamma$}) reaction can also be considered for the element \textbf{lithium-7} ($_{3}{}Li^7$). However, when a neutron is captured by lithium, the radioactive isotope \textbf{beryllium-8} ($_{4}Be^8$) forms. This isotope decays into two helium nuclei, so there is no increase in beryllium concentration due to nuclear transformation through this pathway.
\[
_{3}{}\text{Li}^{7} + n^* \rightarrow (_{3}{}\text{Li}^{8})^* \rightarrow (_{4}{}\text{Be}^{8})^* + \beta^- + \tilde{\nu} + \gamma;
\hspace{1cm}
(_{4}{}\text{Be}^{8})^*\rightarrow _{2}{}\text{He}^{4} + \alpha + \tilde{\nu} + \gamma; \tag{4} \label{eq:4}
\]

An obvious question arises regarding the origin of lithium elements. Given the extremely long half-life of the \boldsymbol{$\beta^-$}-active radioisotope \textbf{beryllium-}10 ($_{4}{}Be^{10}$), it is likely that this isotope can capture an additional quasi-neutron. In this case, along with the reaction (\ref{eq:3}), the following reaction may also occur.
\[
(_{4}{}\text{Be}^{10})^* + n^* \rightarrow (_{4}{}\text{Be}^{11})^* \rightarrow (_{2}{}\text{He}^{7})^* + \alpha + \tilde{\nu} + \gamma, \tag{5} \label{eq:5}
\]
where \textbf{helium-}7 ($_{2}{}He^7$) is a \boldsymbol{$\beta^--$}active isotope, which transforms into the stable isotope lithium-7 ($_{3}{}\text{Li}^{7}$)in a very short period of time ($2.9\cdot10^{-21}$ seconds) ~\cite{20}.
\[
(_{2}{}\text{He}^{7})^* \rightarrow _{3}{}\text{Li}^{7} + \beta^{-} + \tilde{\nu} + \gamma \tag{6} \label{eq:6}
\]
However, the probability of decays described by reactions (\ref{eq:5}) and (\ref{eq:6}) is only 2.9\%. The majority of the decay (97.1\%) proceeds via the following pathway:
\[
(_{4}{}\text{Be}^{10})^* + n^* \rightarrow (_{4}{}\text{Be}^{11})^* \rightarrow _{5}{}\text{B}^{11} + \beta^{-} + \tilde{\nu} + \gamma \tag{7} \label{eq:7}
\]
Such a substantial increase in lithium concentration, as illustrated in Fig~\ref{fig:6}, cannot be accounted for by the decay channels (\ref{eq:5}) and (\ref{eq:6}). Furthermore, the experimental data shows that the quantity of the stable beryllium isotope remains almost unchanged. This indicates the need to explore an alternative pathway for lithium element formation. In water, under normal conditions, certain gases are dissolved, including the inert gas helium, which is also present in significant quantities in the air. Of the two stable helium isotopes, \textbf{helium-}4 ($_{2}{}He^4$) makes up 99.9999\%. Lithium generation based on this helium isotope is possible if the helium nucleus simultaneously captures 2 or 3 quasi-neutrons. The likelihood of two mass-deficient particles merging is notably high, meaning that the formation of a diquasi-neutron ($_{0}{}n^{*2}$) or tri-quasi-neutron ($_{0}{}n^{*3}$) is plausible. Considering this reasoning, the mechanism for such transformations can be represented as follows:
\[
_{2}{}\text{He}^{4} +_{0}{}\text{n}^{*2} \rightarrow (_{2}{}\text{He}^{6})^* \rightarrow _{3}{}\text{Li}^{6} + \beta^{-} + \tilde{\nu} + \gamma \tag{8} \label{eq:8}
\]
\[
_{2}{}\text{He}^{4} + _{0}{}\text{n}^{*3} \rightarrow (_{2}{}\text{He}^{7})^* \rightarrow _{3}{}\text{Li}^{7} + \beta^{-} + \tilde{\nu} + \gamma \tag{9} \label{eq:9}
\]

From the presented experimental measurement results, it is possible to conclude that diquasi-neutrons and tri-quasi-neutrons may form, as per the mechanism described in equation ($\ref{eq:1}$).
        \section{Changes in the Concentration of Stable Isotopes of Certain Chemical Elements Under Acoustic Vibrations in Water}
\begin{figure}[H]
    \centering
    \includegraphics[width=3.2in]{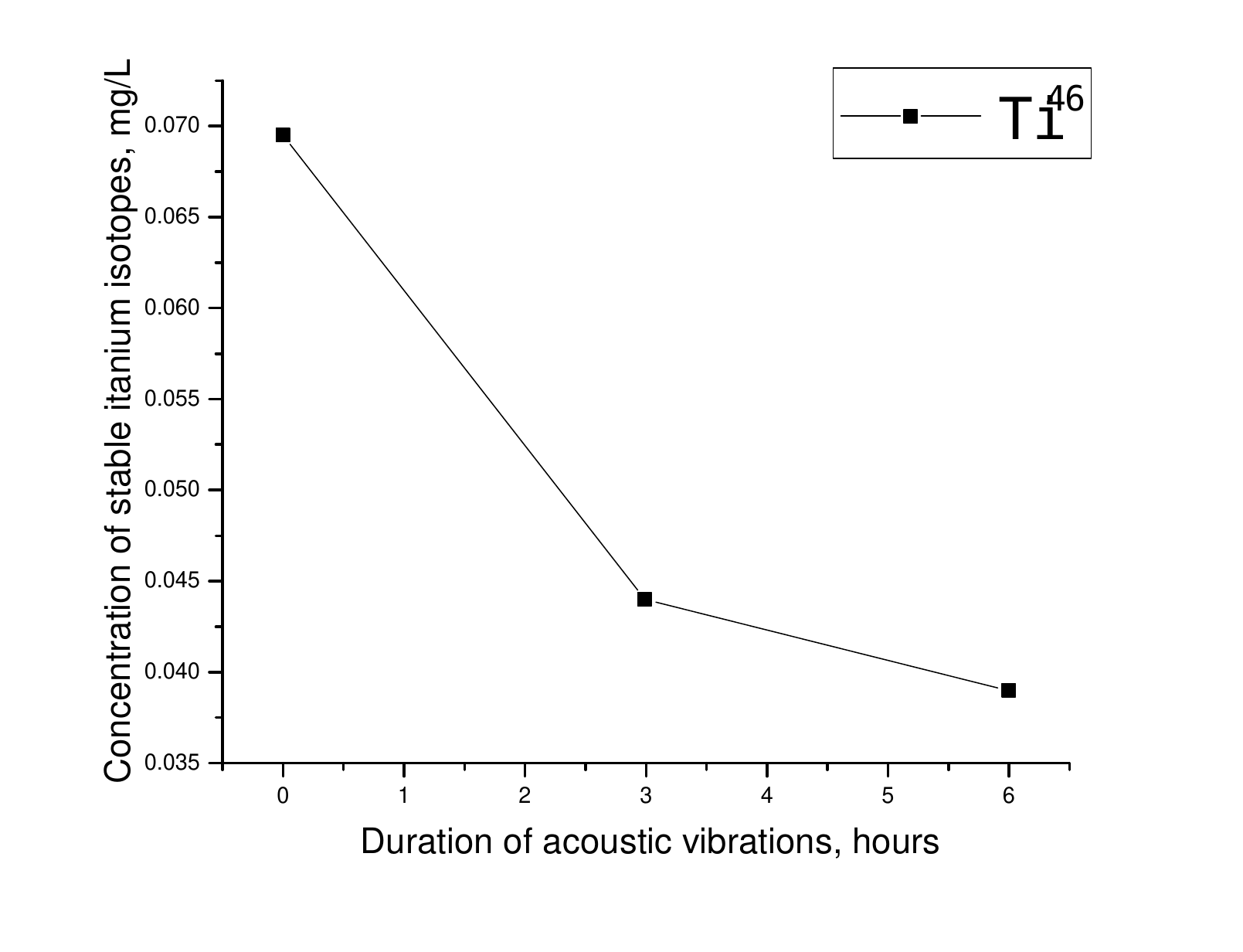} 
    \includegraphics[width=3.2in]{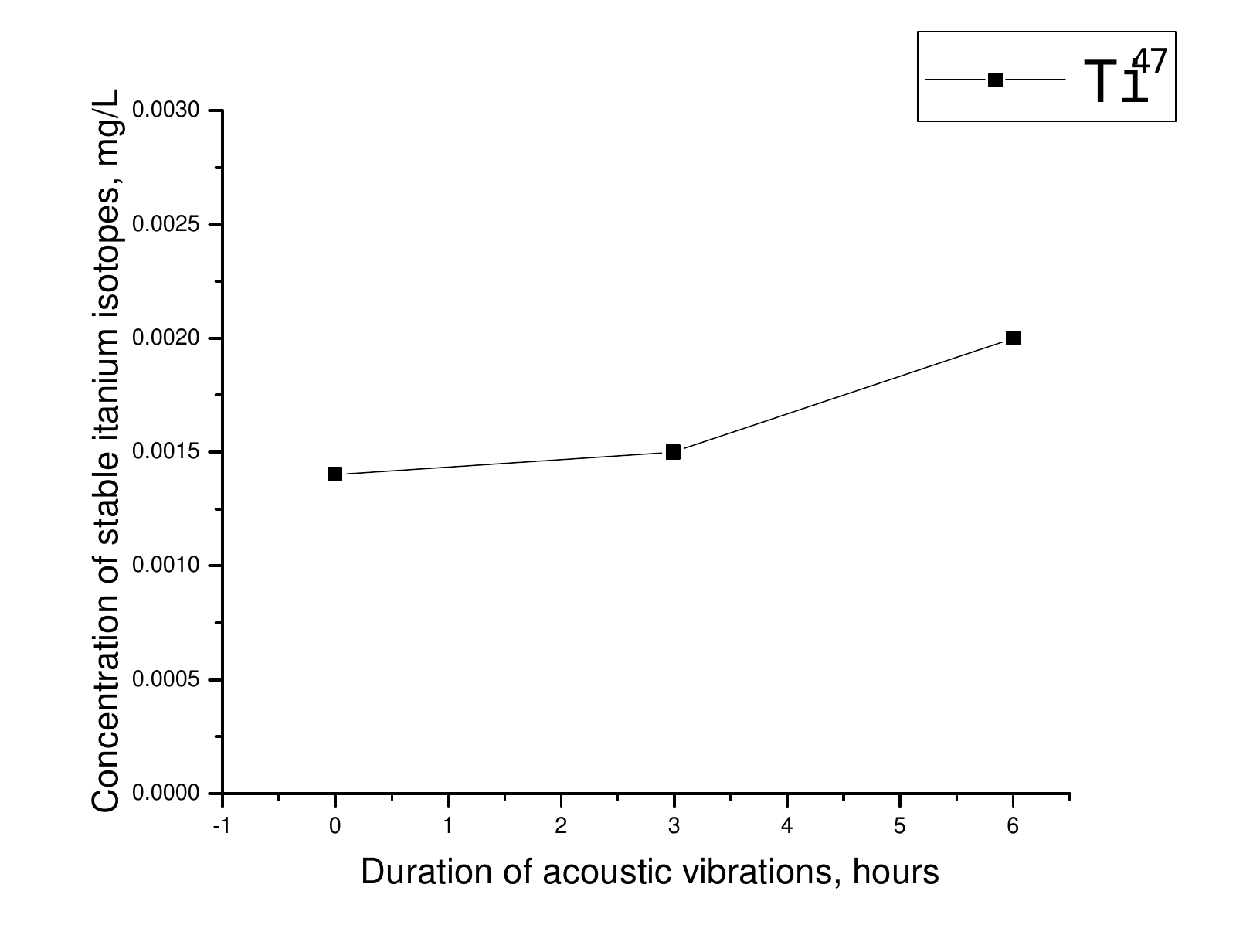} \\
    \includegraphics[width=3.2in]{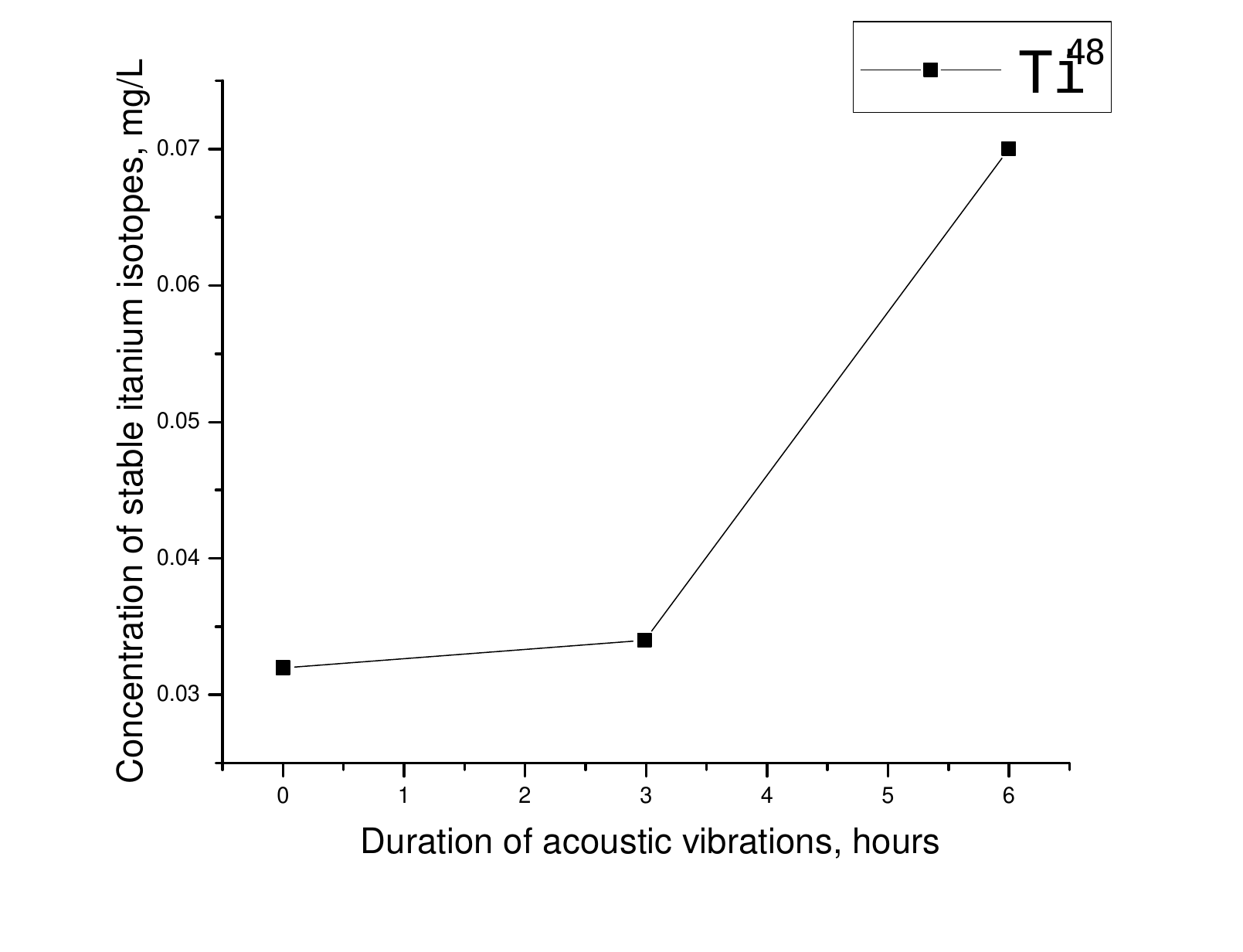} 
    \includegraphics[width=3.2in]{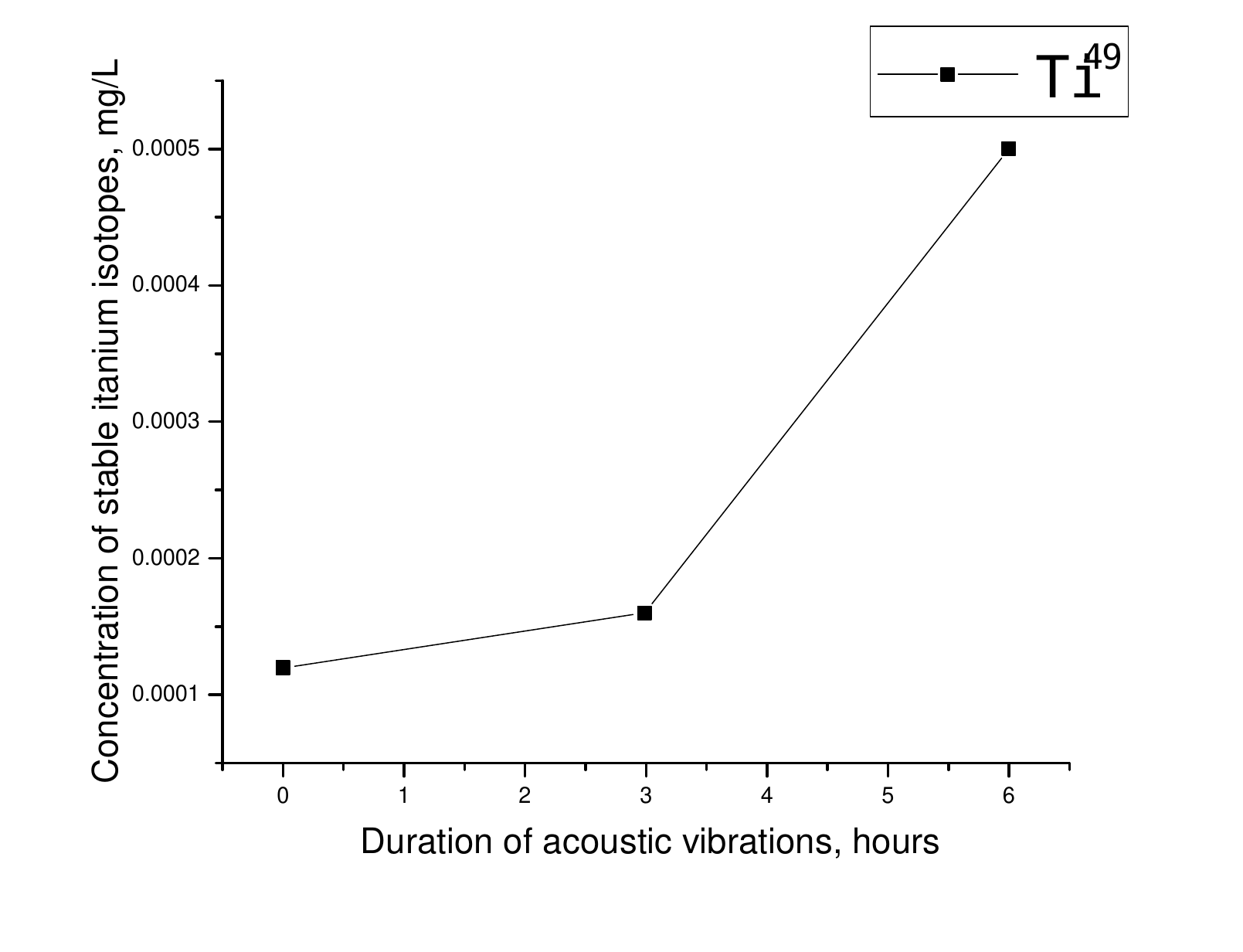} \\
    \includegraphics[width=3.2in]{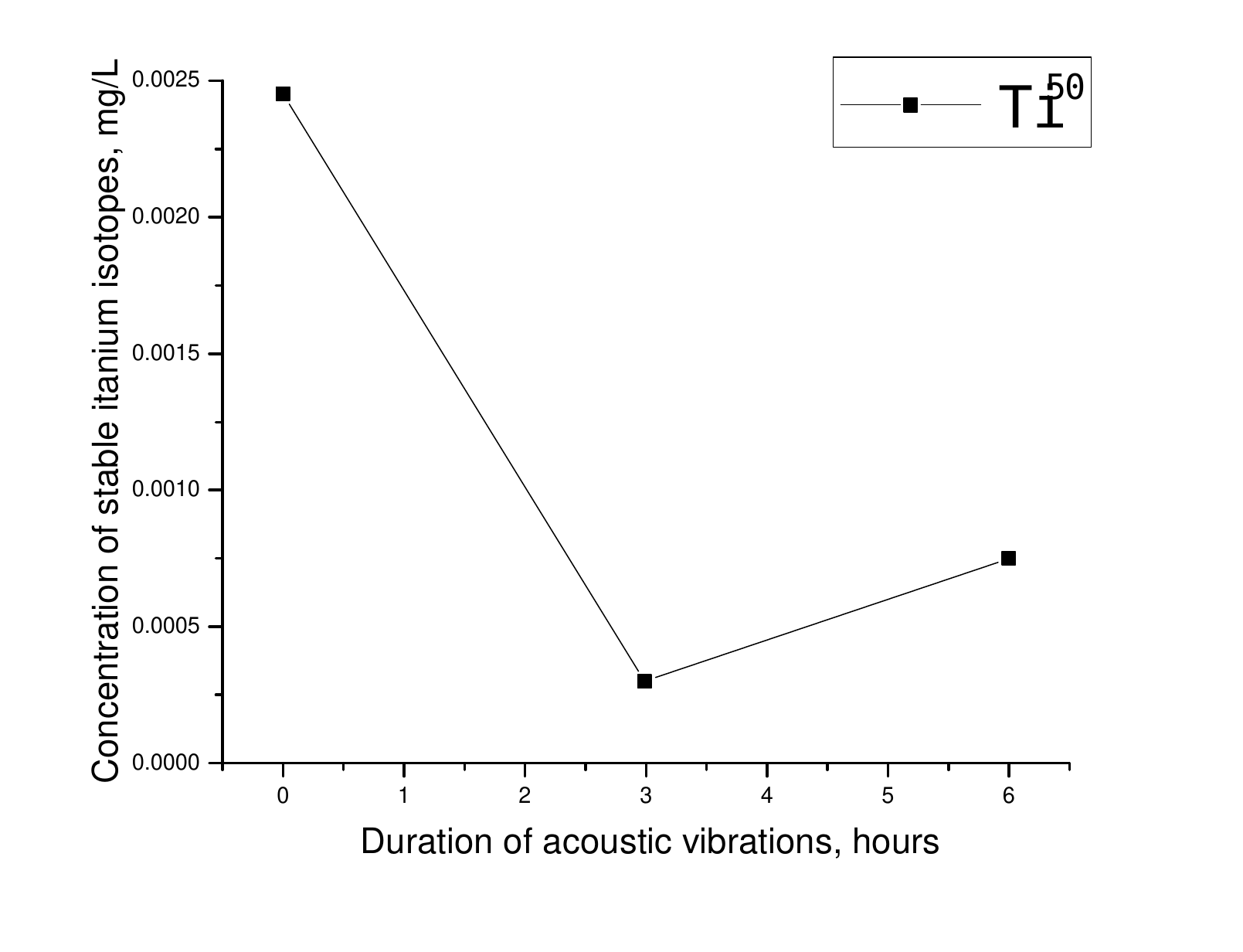} \\
    \caption{Dependence of the concentration (mg/L) of stable titanium isotopes in distilled water on the duration (hours) of the action of acoustic vibrations.}
    \label{fig:7}
\end{figure}
Study~\cite{18} reports on the analysis of distilled water samples exposed to intense mechanical vibrations at a frequency of 25 kHz for durations of three and six hours. To detect changes in the concentrations of selected chemical elements ($Ti, B, Na, Mg, Li$), measurements were conducted using a \textbf{Perkin Elmer ICP MS ELAN 9000} mass spectrometer. Since the acoustic emitter's concentrator is made of titanium, changes in the concentration ratios of stable titanium isotopes may occur when in contact with water. Therefore, the results were compared with control samples that were not subjected to acoustic vibrations but were kept in contact with the titanium concentrator for eight hours.  Measurements revealed significant changes in the concentration of stable titanium isotopes depending on the duration of exposure to acoustic vibrations. 
Fig.~\ref{fig:7} demonstrates graphs depicting the dependence of stable titanium isotope concentrations in distilled water on the duration of exposure to acoustic vibrations. The zero points on the graphs represent the concentrations of the isotopes in the control water samples, which were not exposed to acoustic vibrations but remained in contact with the titanium concentrator for eight hours.
A complex pattern of changes in the concentration of stable titanium isotopes is observed depending on the duration of acoustic vibration exposure from the titanium concentrator on the distilled water sample. For exposure durations up to six hours, there is a decrease in the concentration of the isotope $_{22}{}Ti^{46}$, while increases in the concentrations of the isotopes $_{22}{}Ti^{47}$, $_{22}{}Ti^{48}$, and $_{22}{}Ti^{49}$ are observed. For the isotope $_{22}{}Ti^{50}$, a reduction in concentration is observed during the first three hours of acoustic exposure, followed by a slight increase between three and six hours of exposure. Since these isotopes are stable, the observed concentration changes can be schematically represented as follows:
\begin{equation}
\begin{aligned}
_{22}{}\text{Ti}^{46} + n^* & \rightarrow _{22}{}\text{Ti}^{47};  \quad
_{22}{}\text{Ti}^{47} + n^* & \rightarrow _{22}{}\text{Ti}^{48}; \\
_{22}{}\text{Ti}^{48} + n^* & \rightarrow _{22}{}\text{Ti}^{49}; \quad
_{22}{}\text{Ti}^{49} + n^* & \rightarrow _{22}{}\text{Ti}^{50} 
\end{aligned}
\tag{10} \label{eq:10}
\end{equation}
The reduction in the concentration of the isotope \textbf{titanium}-46 ($_{22}{}Ti^{46}$) is likely caused either by the absence of the element \textbf{scandium-}45 ($_{21}{}Sc^{45}$) in the water or by the relatively long half-life (83.8 days)~\cite{20} of the \boldsymbol{$\beta^-$}-decaying radioactive isotope \textbf{scandium-}46 $(_{21}{}Sc^{46})^*$, which exceeds the duration of the acoustic exposure to the water.

\begin{figure}
    \centering
    \includegraphics[width=3.2in]{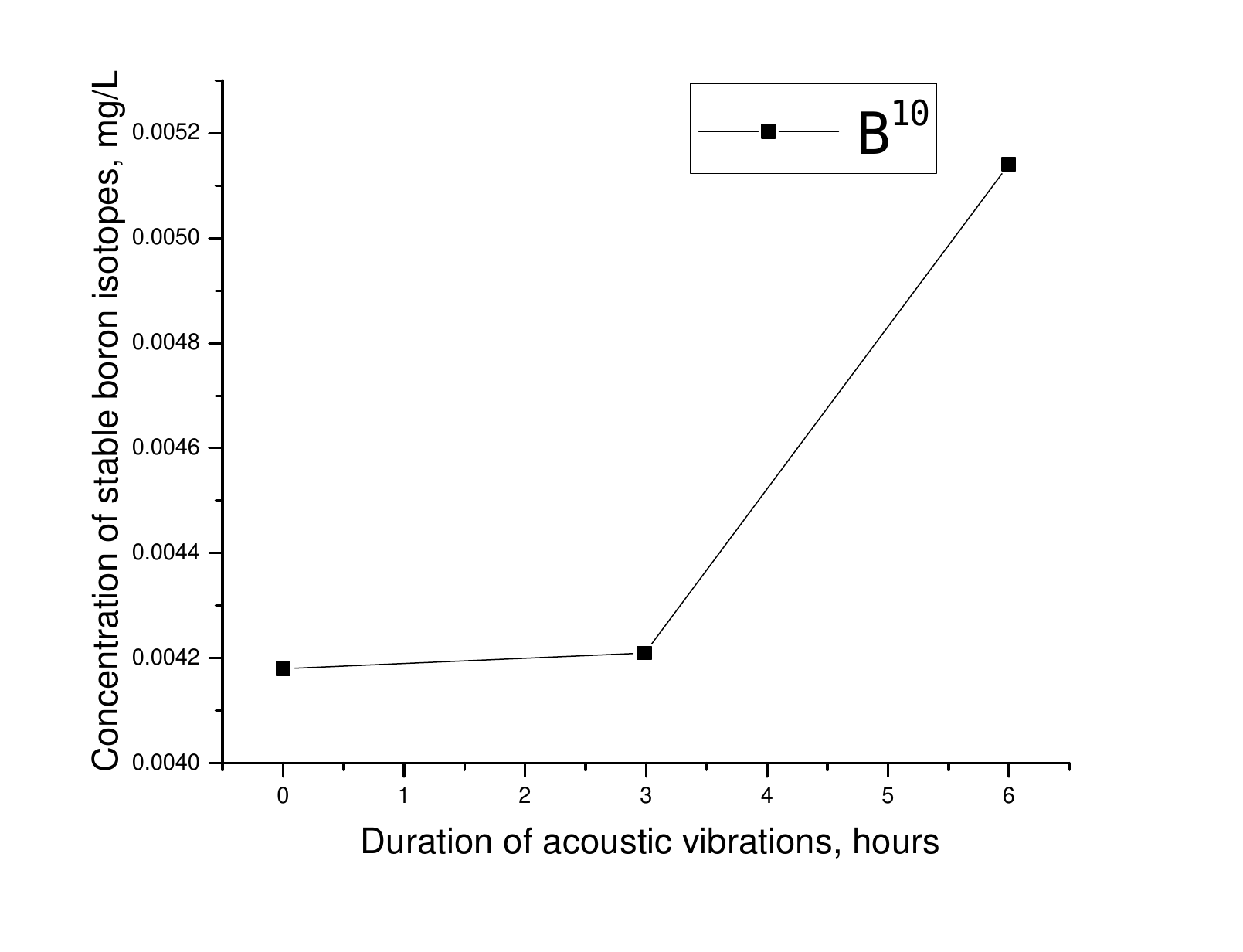} 
    \includegraphics[width=3.2in]{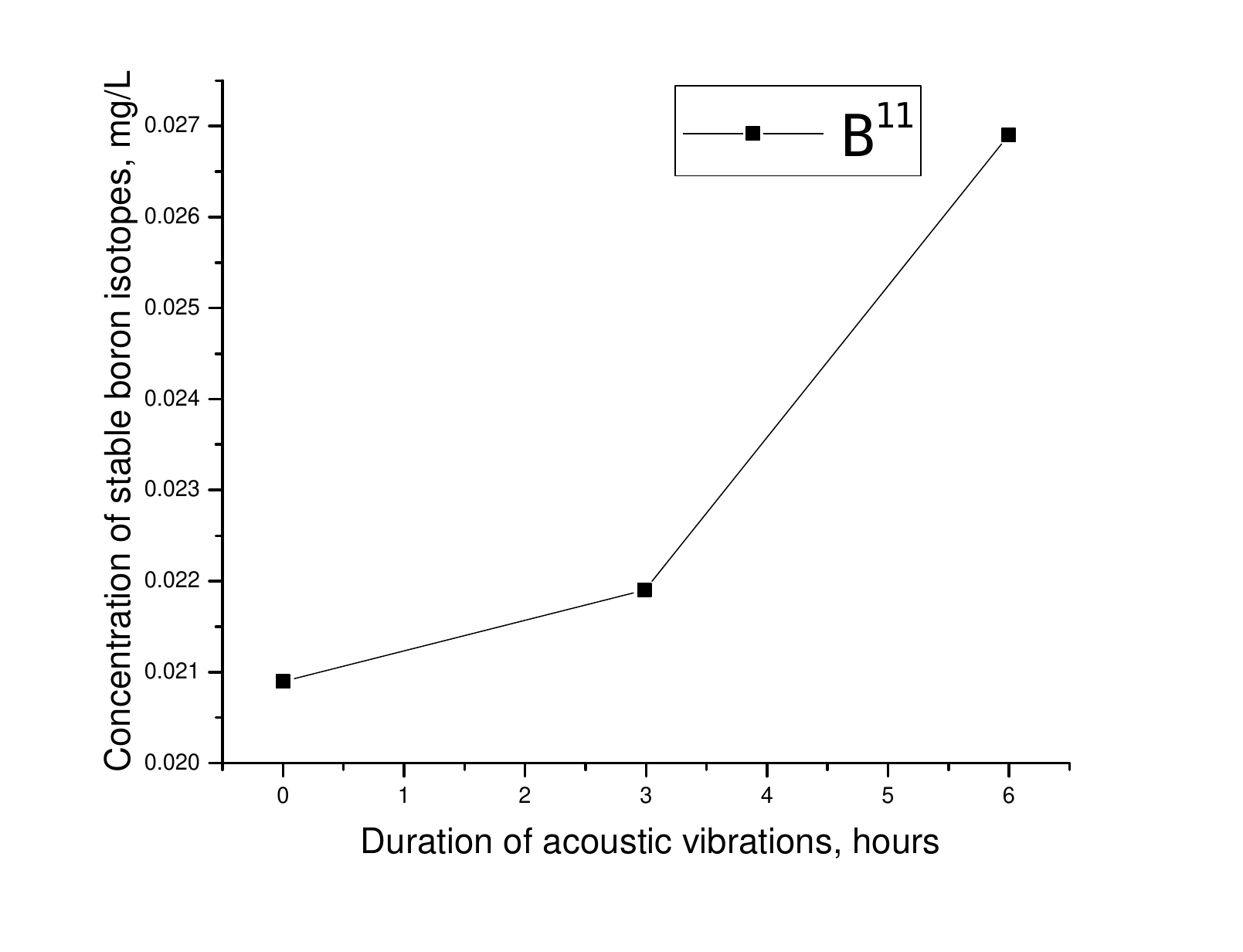} \\
    \caption{Dependence of the concentration (mg/L) of stable boron isotopes in distilled water on the duration (hours) of the action of acoustic vibrations.}
    \label{fig:8}
\end{figure}

Fig~\ref{fig:8} illustrates the changes in concentration, (mg/L), of boron isotopes as a function of the exposure time, (hours), to acoustic vibrations in water. As indicated by the graphs, the magnitude of the change in the concentration of stable boron isotopes is small and can be explained by the following mechanism:
\begin{equation}
\begin{aligned}
_{4}{}\text{Be}^{9} + n^* & \rightarrow (_{4}{}\text{Be}^{10})^* \rightarrow _{5}{}\text{B}^{10} + \beta^- + \tilde{\nu} + \gamma; \quad
 & _{5}{}\text{B}^{10} + n^*  \rightarrow _{5}{}\text{B}^{11}
\end{aligned}
\tag{11} \label{eq:11}
\end{equation}

% \begin{equation}
% \begin{array}{rl}
% _{4}{}\text{Be}^{9} + n^* & \rightarrow (_{4}{}\text{Be}^{10})^* \rightarrow _{5}{}\text{B}^{10} + \beta^- + \tilde{\nu} + \gamma;\\
%  & _{5}{}\text{B}^{10} + n^*  \rightarrow _{5}{}\text{B}^{11}
% \end{array}
% \tag{11} \label{eq:11}
% \end{equation}

\begin{figure}
    \centering
    \includegraphics[width=3.9in]{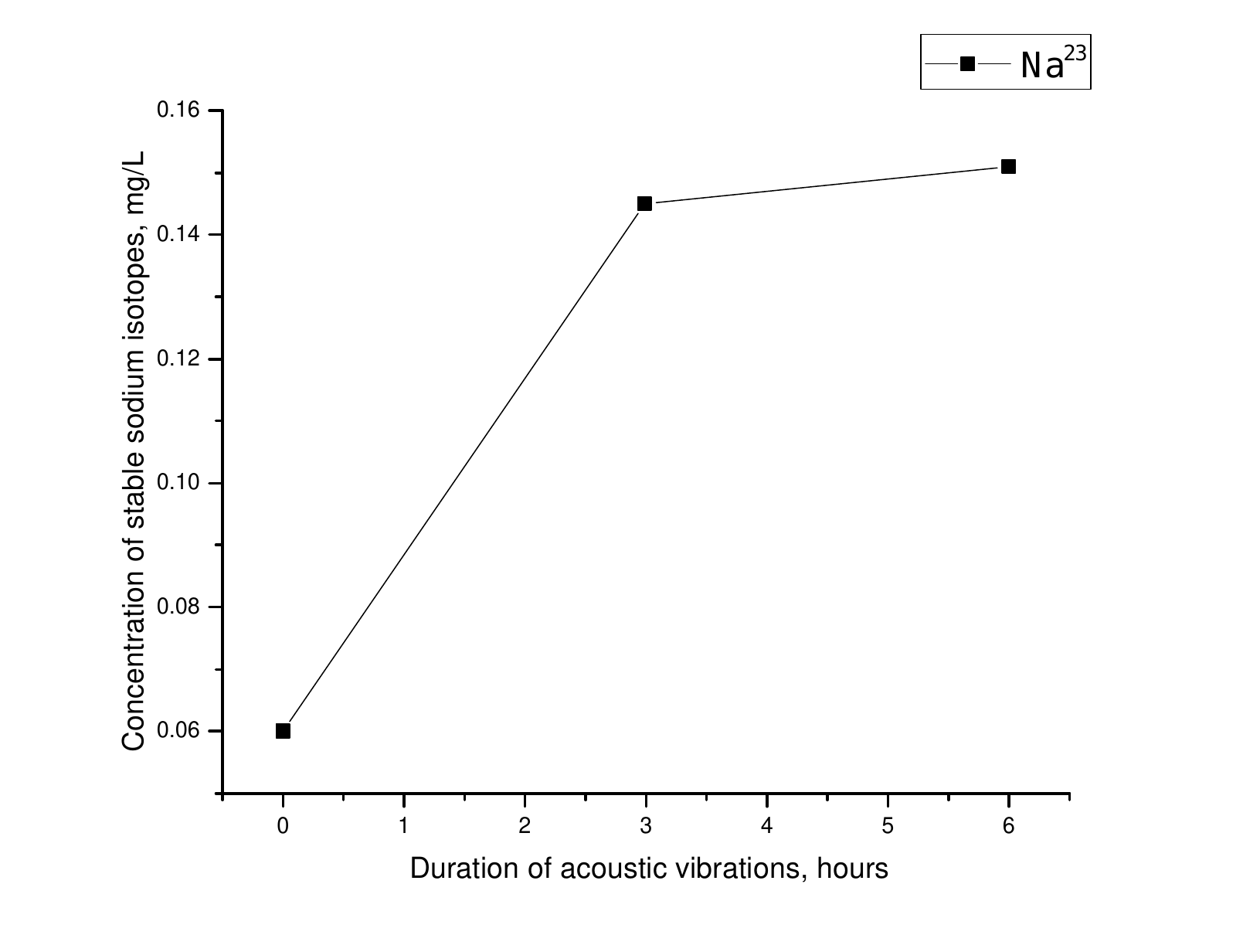} 
    \caption{Dependence of the concentration (mg/L) of stable sodium isotopes in distilled water on the duration (hours) of the action of acoustic vibrations.}
    \label{fig:9}
\end{figure}

Fig.~\ref{fig:9} illustrates a significant increase in the concentration of the stable sodium isotope \textbf{sodium-23} $(_{11}{}Na^{23})$. This increase can be attributed to the presence of the stable neon isotope \textbf{neon-22} $(_{10}{}Ne^{22})$ in the water, as described by the following mechanism:
\[
_{10}{}\text{Ne}^{22} + n^* \rightarrow (_{10}{}\text{Ne}^{23})^* \rightarrow _{11}{}\text{Na}^{23} + \beta^- + \tilde{\nu} + \gamma
\tag{12} \label{eq:12}
\]
\newpage
Under the influence of sharp mechanical oscillations with an acoustic frequency, changes in the concentration of stable magnesium isotopes in water are also observed.

\begin{figure}[H]
    \centering
    \includegraphics[width=3.1in]{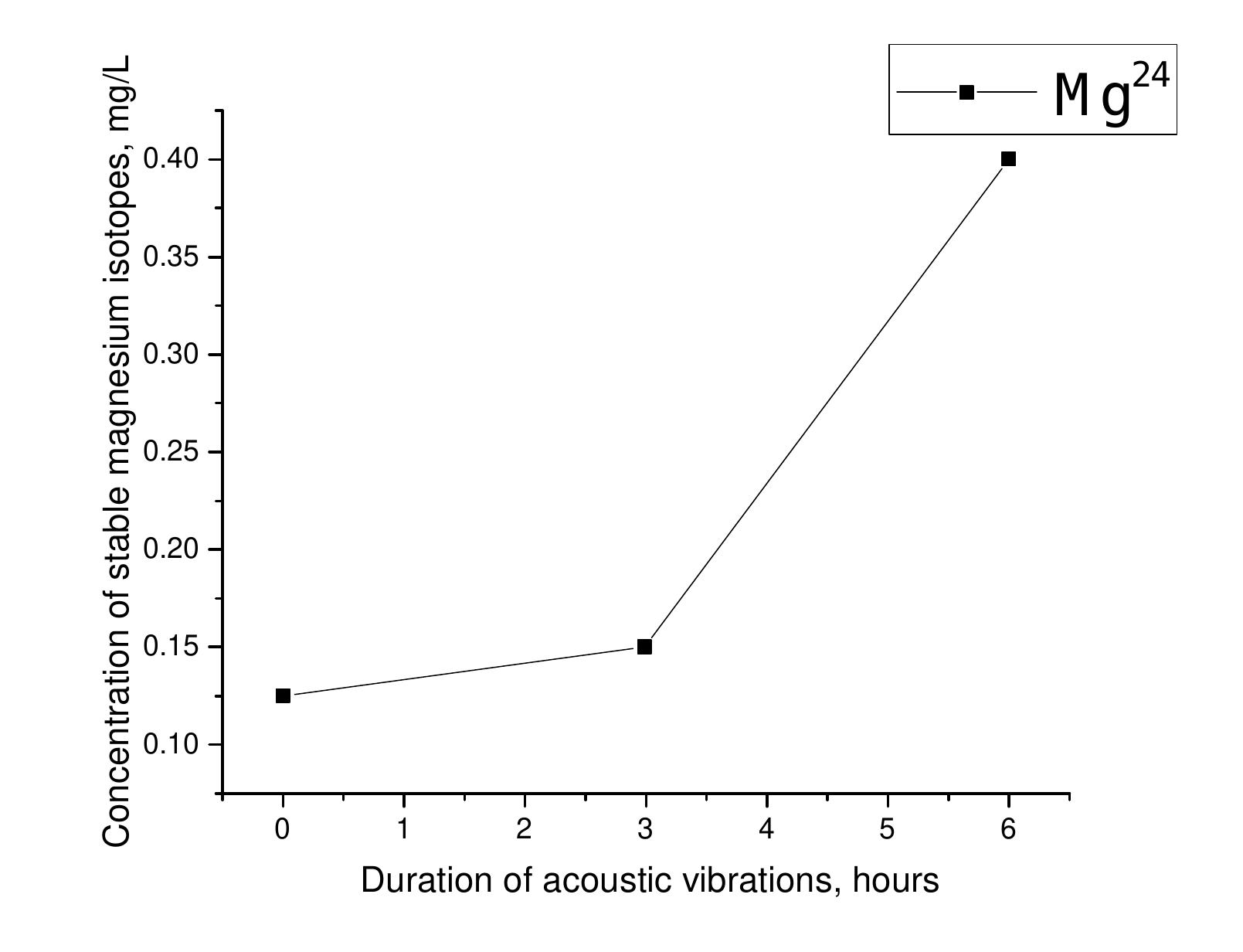} 
    \includegraphics[width=3.1in]{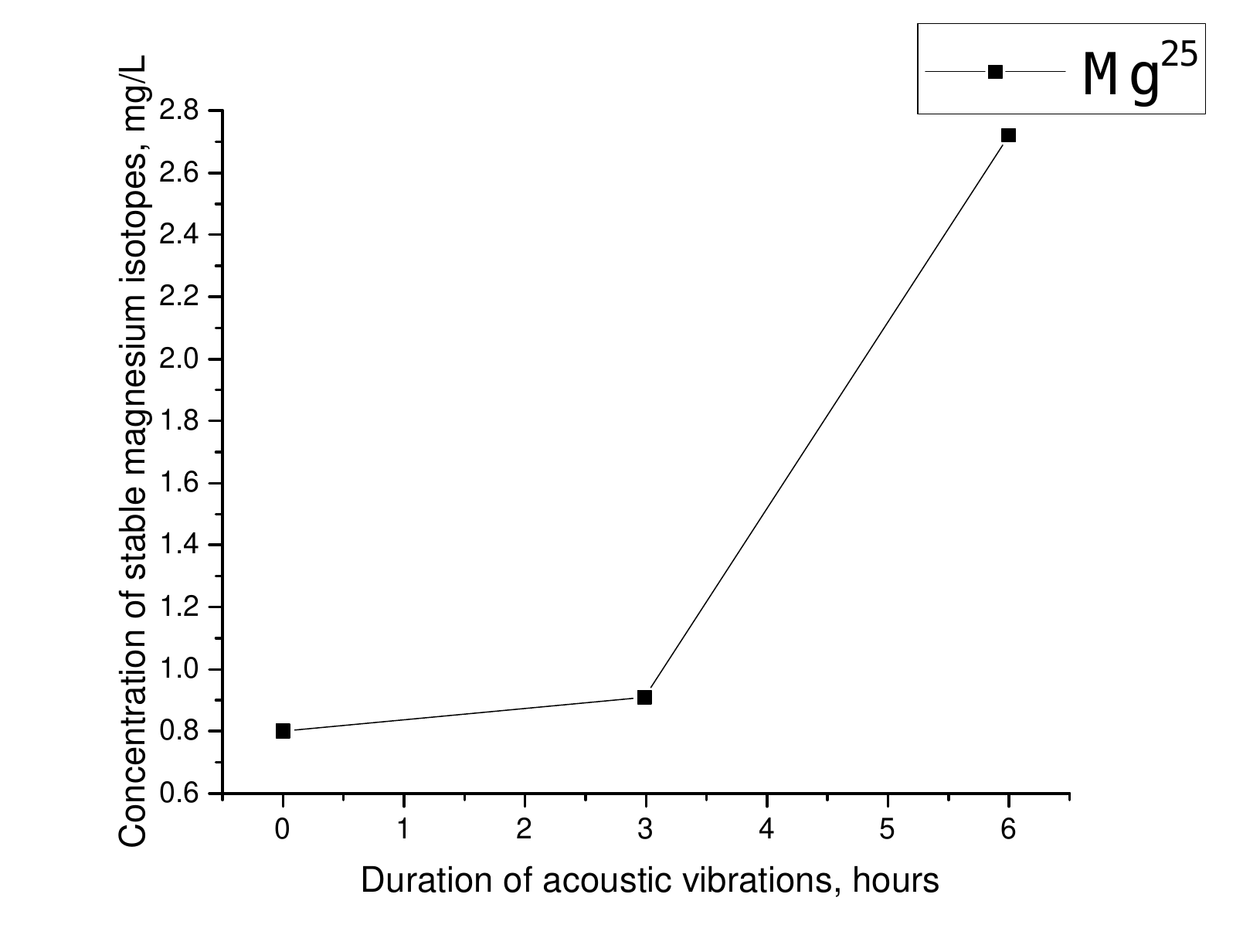} 
    \includegraphics[width=3.1in]{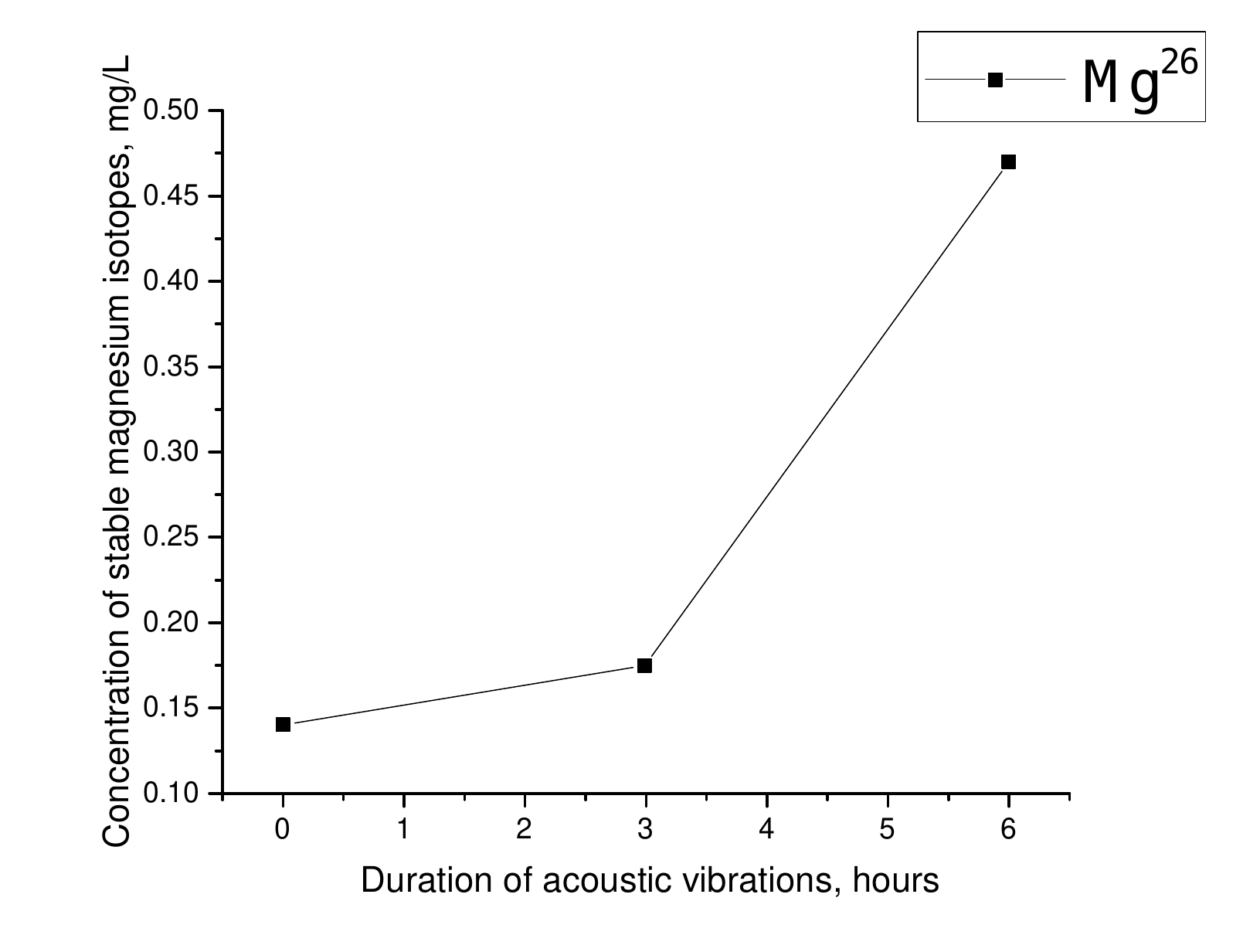} 
    \caption{Dependence of the concentration (mg/L) of stable magnesium isotopes in distilled water on the duration (hours) of the action of acoustic vibrations.}
    \label{fig:10}
\end{figure}

Fig.~\ref{fig:10} illustrates the relationship between the concentrations of three stable magnesium isotopes and the duration of exposure to acoustic oscillations.

A notable feature of these changes is the identical pattern observed in the concentration of all stable magnesium isotopes under the influence of ultrasonic oscillations. Similarly to the explanations provided above, the increase in magnesium isotope concentrations can be attributed to the probable capture of quasi-neutrons by stable sodium isotopes according to the following scheme:
\[
_{11}{}\text{Na}^{23} + n^* \rightarrow (_{11}{}\text{Na}^{24})^* \rightarrow _{12}{}\text{Mg}^{24} + \beta^- + \tilde{\nu} + \gamma
\tag{13} \label{eq:13}
\]
\begin{equation}
\begin{array}{rl}
_{12}{}\text{Mg}^{24} + n^* & \rightarrow _{12}{}\text{Mg}^{25};\\
_{12}{}\text{Mg}^{25} + n^* & \rightarrow _{12}{}\text{Mg}^{26};
\end{array}
\tag{14} \label{eq:14}
\end{equation}

\begin{figure}
    \centering
    \includegraphics[width=3.2in]{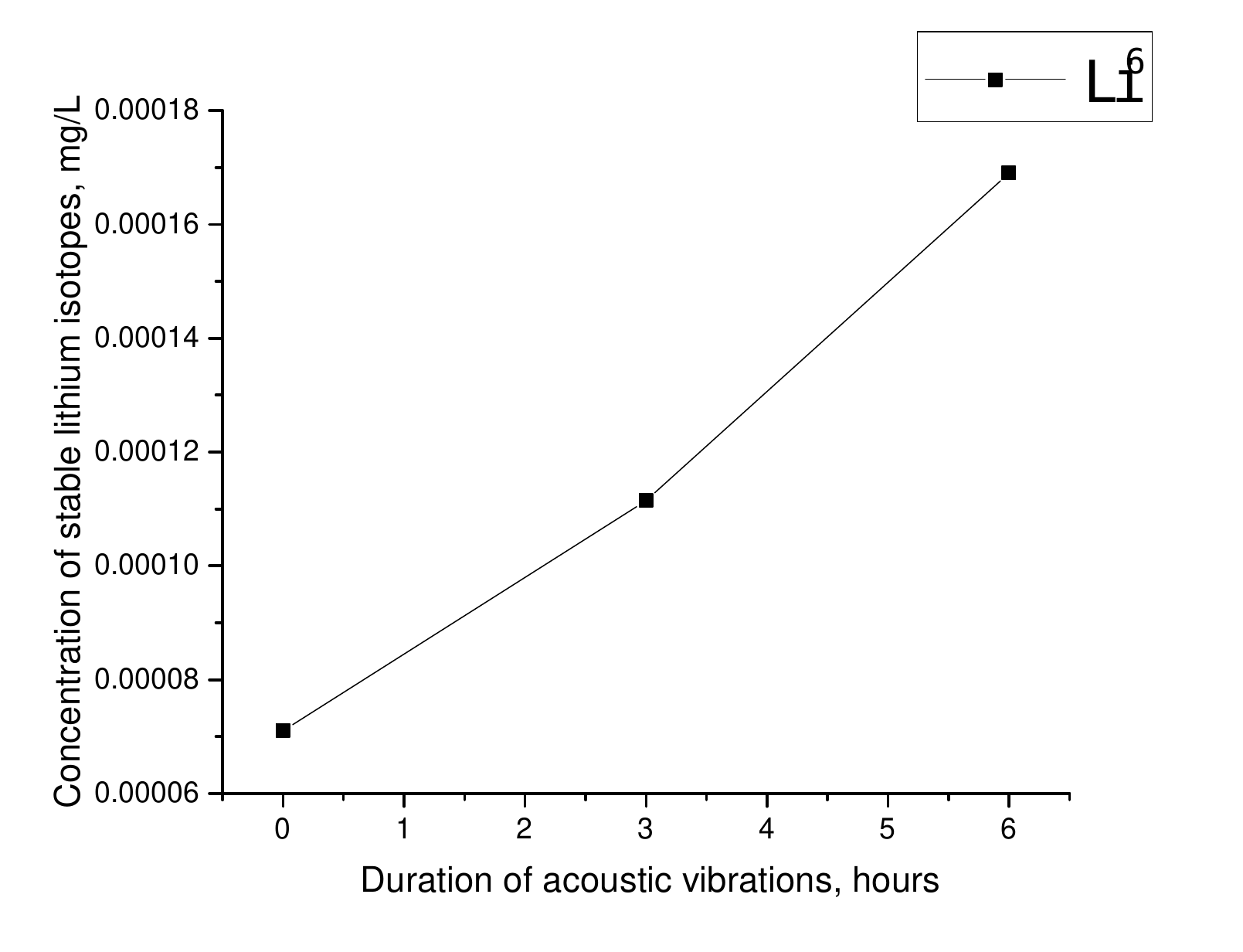} 
    \includegraphics[width=3.2in]{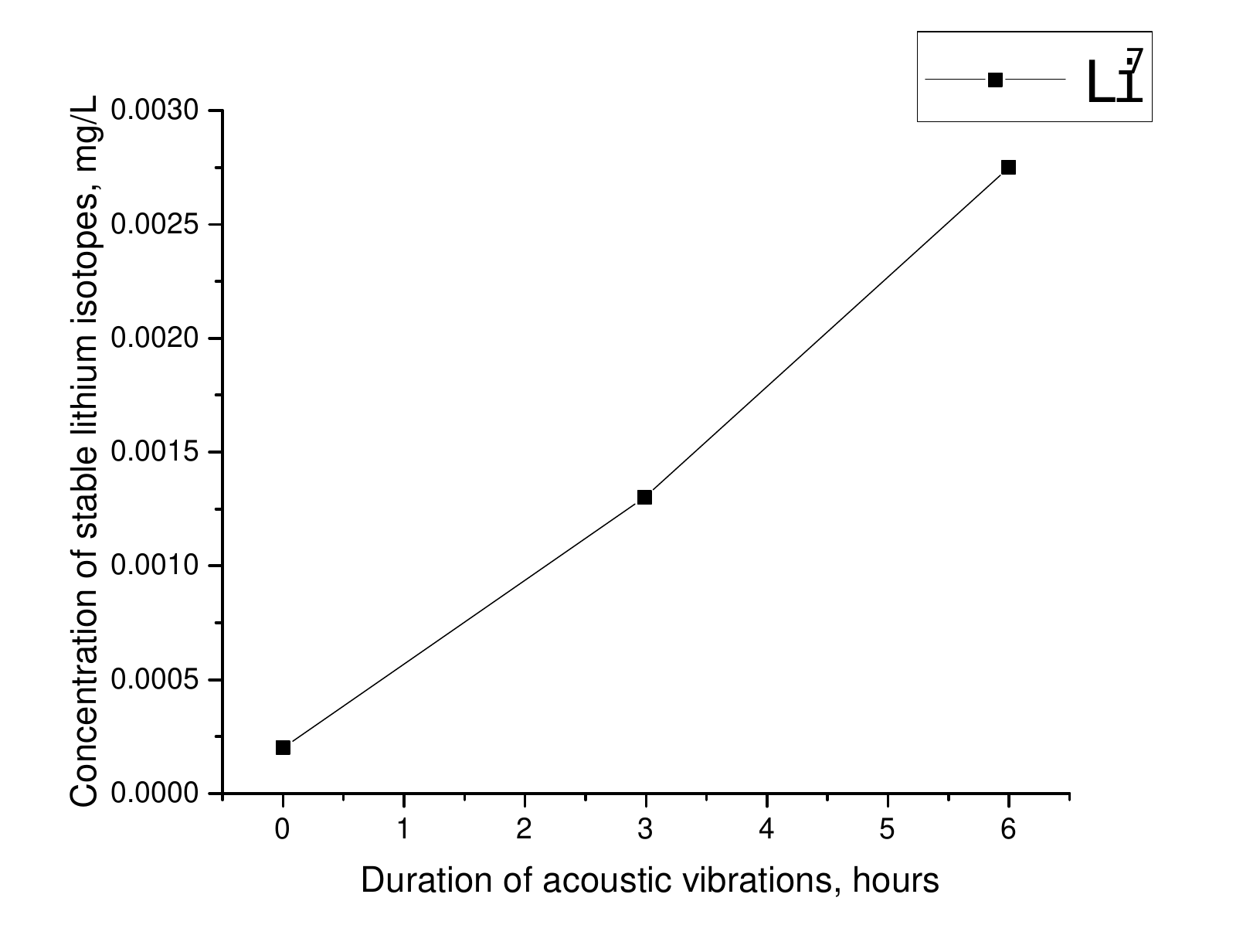} 
    \caption{Dependence of the concentration (mg/L) of stable lithium isotopes in distilled water on the duration (hours) of the action of acoustic vibrations.}
    \label{fig:11}
\end{figure}

Fig.~\ref{fig:11} illustrates the relationship between the concentration  (mg/l) of stable lithium isotopes in distilled water and the duration (hours) of exposure to acoustic vibrations, indicating a significant increase in the concentrations of the two stable isotopes \textbf{lithium-6} ($_{3}{}Li^6$) and \textbf{lithium-7} ($_{3}{}Li^7$). 

Determining the source of lithium element formation is not straightforward. This complexity arises from the significant difference between the atomic weight of the element preceding lithium, ($_{2}{}He^4$), and that of lithium. However, as previously mentioned, based on the concept of the formation of diquasineutrons and triquasineutrons, it becomes possible to propose a channel for the formation of stable lithium isotopes according to schemes (\ref{eq:8}) and (\ref{eq:9}). The study~\cite{19} presents the results of experimental investigations into the implementation of acoustically induced nuclear processes in heavy water under normal conditions. In experiments with both ordinary water~\cite{17} and heavy water, under mechanical action with steep fronts and acoustic repetition frequency, \boldsymbol{$\gamma$}-radiation was recorded. To confirm the presence of nuclear transmutations in heavy water, mass spectrometric measurements of impurity concentrations in samples were conducted before and after exposure to sharp mechanical impulses. Comparative characteristics of the measurement results in ordinary water, replicating the experimental conditions, are also presented. Significant changes in impurity concentrations in the liquids, both quantitatively and qualitatively, were observed. Based on the understanding of the structure of double electrical layers that arise upon contact between metal and heavy water, a mechanism for the formation of quasineutron complexes is proposed. These complexes participate in nuclear transmutations, resulting in changes in the concentration of impurities in heavy water. These complexes participate in nuclear transmutations, resulting in changes in the concentration of impurities in heavy water. Thus, such complex changes in the isotopic composition of certain chemical elements may be related to the conditions for quasineutron formation and the probability of chemical element nuclei being near the generated quasineutron.
          \section{Conclusions}
Experimental results are presented that demonstrate the phenomenon of acoustically induced nuclear processes in water, confirmed by direct measurements of radiation emissions and the formation of new elements.  The emergence of these elements cannot be explained by chemical processes. These phenomena were observed in regular water when subjected to acoustic impulses with steep fronts (less than 50$\mu\text{s}$). There is no need for cavitation and sonoluminescence phenomena to observe these processes. It was shown that the concentration of chemical elements in water changes under abrupt mechanical impacts. The complex nature of the influence of mechanical vibrations on the concentration of stable isotopes of the chemical elements $Ti, B, Na, Mg,$ and $Li$ in water is also demonstrated.

The complex changes in element concentrations can be explained by differences in the cross-sections of nuclear reactions like ($n$, \boldsymbol{$\gamma$}), the half-life of the radioactive isotopes formed after quasineutron capture, and the proximity of elements to the titanium concentrator, near which quasineutrons are generated. The possible mechanism for the formation of quasineutrons and their complexes, which participate in nuclear processes, is explained by the presence of a double electric layer formed at the interface of the media. The erosion of metal surfaces during cavitation in water is explained by the production of fluorine and, consequently, the formation of the aggressive hydrofluoric acid (HF) molecules. A mechanism for the occurrence of sonoluminescence has been proposed, explaining the origin of the continuous optical emission spectrum observed during the process.
        \bibliography{bibliography}
\end{document}